\documentclass[]{elsarticle}
\usepackage{graphicx}
\usepackage{amsmath}
\usepackage[utf8]{inputenc}
\usepackage{comment}
\usepackage{ragged2e}
\usepackage{color}
\usepackage{listings,multicol}
\usepackage[linesnumbered,algoruled,boxed,lined]{algorithm2e}
\usepackage{tabularx}
\usepackage{array}
\newcolumntype{L}{>{\centering\arraybackslash}m{6cm}}
\usepackage{url}
\usepackage{breakurl}
\providecommand{\href}[2]{#2}
\usepackage{booktabs}
\usepackage{float}
\usepackage{placeins}
\usepackage{needspace}
\usepackage{tikz}
\usetikzlibrary{positioning,arrows.meta,fit,backgrounds,calc,decorations.pathreplacing,shapes.geometric}

\lstdefinestyle{myListingStyle}
    {
        mathescape,
        basicstyle = \small\ttfamily,
        breaklines = true,
        keepspaces=true,
        showstringspaces=false,
        columns=fullflexible,
        emph={filetype_}, emphstyle=\textcolor{blue}
    }

\AtBeginDocument{\raggedbottom}

\AtBeginDocument{%
  \providecommand\BibTeX{{%
    \normalfont B\kern-0.5em{\scshape i\kern-0.25em b}\kern-0.8em\TeX}}}

\begin{document}
\author[lsu,cyber]{Karley~Waguespack}
\ead{kwagu14@lsu.edu}

\author[lsu,cyber]{Samuel~Goodwin}
\ead{sgood11@lsu.edu}

\author[lsu,cyber]{George~Hendrick}
\ead{ghendr4@lsu.edu}

\author[lsu,cyber]{Samuel~Hildebrand}
\ead{shilde2@lsu.edu}

\author[lsu,cyber]{Thomas~Landaiche~III}
\ead{tlanda1@lsu.edu}

\author[lsu,cyber]{Mingyang~Li}
\ead{myi5@lsu.edu}

\author[lsu,cyber]{Joshua~McCain}
\ead{jmcca51@lsu.edu}

\author[lsu,cyber]{Tyler~Saizan}
\ead{tsaiza2@lsu.edu}

\author[lsu,cyber]{Jacob~Tucker}
\ead{jtuck49@lsu.edu}

\author[lsu,cct,cyber]{James~M.~Ghawaly}
\ead{jghawaly@lsu.edu}

\author[lsu,cct,cyber]{Golden~G.~Richard~III\corref{cor1}}
\ead{golden@cct.lsu.edu}
\cortext[cor1]{Corresponding author}

\address[lsu]{Division of Computer Science and Engineering, Louisiana State University}

\address[cct]{Center for Computation and Technology, Louisiana State University}

\address[cyber]{LSU Cyber Center, Louisiana State University}

\title{Scalpel3: A High-Performance Data Carving Architecture for Recovery of Fragmented Files}

\begin{abstract}

  File carving recovers files from raw storage without filesystem metadata, a key capability in
  digital forensics, data recovery, and digital exploration. Existing tools recover contiguous
  files effectively, but, to our knowledge, no publicly available, format-agnostic,
  high-performance framework exists in which researchers can develop and deploy new fragmented
  recovery strategies. Scalpel3 fills this gap with a massively threaded architecture for
  contiguous and fragmented recovery. Researchers need only write single-threaded validation and
  reassembly code for a new file type; Scalpel3 supplies worker scheduling, synchronization,
  checkpointing, and I/O. This separation allows new recovery methods to be added without modifying
  the backend infrastructure. The architecture also integrates the ONNX Runtime, allowing learned
  models to be used within validators and recovery strategies. Operational features include
  interactive human-in-the-loop control, block
  deduplication, persistent restart checkpoints, incremental output, and a FUSE filesystem for
  hybrid workflows.

  We evaluate Scalpel3 on a mixed corpus of more than 80{,}000 files under contiguous recovery and
  three controlled fragmentation scenarios: gaps, out-of-order block placement, and both together. Results show
  fast and accurate contiguous recovery and demonstrate that Scalpel3's massively threaded
  architecture makes validated fragmented results available substantially earlier than
  single-threaded execution. Furthermore, strategies tailored to individual file types maintain high overall
  accuracy across increasingly difficult layouts. Together, these results demonstrate that
  Scalpel3 provides a practical foundation for developing and deploying fragmented recovery
  strategies at scale.

\end{abstract}


\maketitle

\section{Introduction}

File carving is the process of recovering files from raw disk images or other media without relying on filesystem metadata. This technique is important in data recovery tasks where filesystem metadata is missing or intentionally destroyed, such as after file deletion or drive reformatting. Because it recovers data from file contents alone, carving can also recover files from damaged storage media, where filesystem metadata may be unreadable. File carving also supports ``digital exploration'': examining a storage device of unknown origin without prior knowledge of its contents or a predefined investigative target.

\begin{figure}[t]
  \centering
  \resizebox{\textwidth}{!}{%
  \begin{tikzpicture}[
    font=\sffamily,
    blk/.style={draw, minimum width=0.95cm, minimum height=0.9cm, inner sep=1pt,
                font=\sffamily\small, align=center},
    f5/.style={blk, fill=cyan!22},
    ot/.style={blk, fill=black!12},
    fr/.style={blk, fill=white},
    ttl/.style={font=\sffamily\small},
    sub/.style={font=\sffamily\footnotesize, text=black!75, align=center},
    arr/.style={-{Stealth[length=3mm]}, very thick, color=purple!70!black},
    brc/.style={decorate, decoration={brace, amplitude=5pt, mirror, raise=2pt},
                color=black!55},
  ]
    \node[ot] (a1) {};
    \node[f5, right=0pt of a1]  (a2)  {\textbf{H}};
    \node[f5, right=0pt of a2]  (a3)  {};
    \node[f5, right=0pt of a3]  (a4)  {};
    \node[ot, right=0pt of a4]  (a5)  {};
    \node[ot, right=0pt of a5]  (a6)  {};
    \node[fr, right=0pt of a6]  (a7)  {\scriptsize free};
    \node[f5, right=0pt of a7]  (a8)  {};
    \node[f5, right=0pt of a8]  (a9)  {};
    \node[f5, right=0pt of a9]  (a10) {\textbf{F}};
    \node[ot, right=0pt of a10] (a11) {};
    \node[ot, right=0pt of a11] (a12) {};

    \node[ttl, anchor=south] at ($(a1.north west)!0.5!(a12.north east)+(0,0.2)$)
      {Disk image as seen by the carver: no filesystem metadata};

    \draw[brc] (a2.south west) -- (a4.south east)
      node[midway, below=9pt, sub]{target file: fragment 1};
    \draw[brc] (a8.south west) -- (a10.south east)
      node[midway, below=9pt, sub]{fragment 2};

    \coordinate (cx) at ($(a1.south west)!0.5!(a12.south east)$);
    \node[f5, anchor=north west] (b1) at ($(cx)+(-2.85,-1.95)$) {\textbf{H}};
    \node[f5, right=0pt of b1] (b2) {};
    \node[f5, right=0pt of b2] (b3) {};
    \node[f5, right=0pt of b3] (b4) {};
    \node[f5, right=0pt of b4] (b5) {};
    \node[f5, right=0pt of b5] (b6) {\textbf{F}};

    \draw[arr] ($(cx)+(0,-1.05)$) -- ($(cx)+(0,-1.75)$);
    \node[sub, anchor=west] at ($(cx)+(0.3,-1.4)$)
      {header/footer search,\\block validation, reassembly};

    \node[ttl, anchor=north] at ($(b1.south west)!0.5!(b6.south east)+(0,-0.16)$)
      {Recovered file: fragments selected and joined};

    \begin{scope}[shift={($(a1.south west)+(0.4,-4.0)$)}]
      \node[f5, minimum width=0.5cm, minimum height=0.5cm] (l1) {};
      \node[sub, right=3pt of l1, anchor=west] (l1t) {Target file};
      \node[ot, minimum width=0.5cm, minimum height=0.5cm, right=1.0cm of l1t.east, anchor=west] (l2) {};
      \node[sub, right=3pt of l2, anchor=west] (l2t) {Other file data};
      \node[fr, minimum width=0.5cm, minimum height=0.5cm, right=1.0cm of l2t.east, anchor=west] (l3) {};
      \node[sub, right=3pt of l3, anchor=west] {Free space};
    \end{scope}
  \end{tikzpicture}%
  }
  \caption{The fragmented file carving problem. The carver sees only a raw sequence of blocks with no filesystem metadata. The target file is split into two fragments separated by unrelated data and free space; header and footer signatures (\textbf{H}, \textbf{F}) bound the candidate, and block validation and reassembly recover and join its blocks into the original file.}
  \label{fig:basic-frag}
\end{figure}
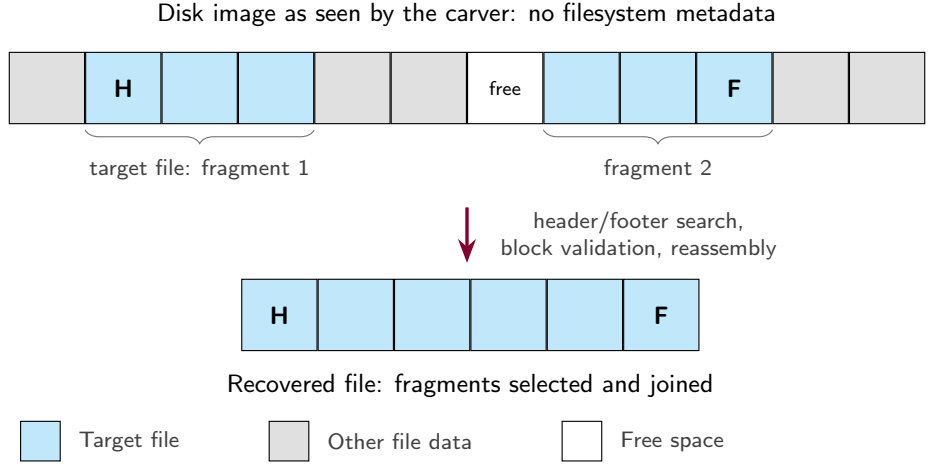

Traditional file carvers operate by scanning for header and footer signatures to identify file boundaries, performing consistency checks, and copying all data between these markers. This technique assumes that files are stored contiguously, an assumption that holds for many files in practice, although in many cases files are not stored contiguously~\cite{VanDerMeer2021Fragmentation}.  Files can become fragmented as they grow or are updated, particularly on volumes with limited contiguous free space. When a file is fragmented, its data is split into multiple noncontiguous regions, interleaved with unrelated data and free space. Figure~\ref{fig:basic-frag} illustrates the resulting recovery problem: a carver sees only a raw sequence of blocks with no filesystem metadata, and must locate a fragmented file's blocks, distinguish them from unrelated data, and reassemble them in the correct order. Traditional file carving tools, which simply copy the contiguous span between a header and footer, are not equipped to recover such files.

Device damage can also leave only partial file fragments recoverable. Traditional carving may recover contiguous remnants, but it cannot reconstruct a file whose surviving blocks are noncontiguous or out of order. Even with manual intervention, reconstructing a file from disordered fragments remains challenging.  Worse, naively trying every possible selection and ordering of candidate blocks creates a vast combinatorial search space that quickly becomes infeasible to handle.

This paper introduces Scalpel3, a flexible and extensible open source file carving framework~\cite{Scalpel3Software}. While earlier versions of Scalpel focused on high-speed contiguous carving on computers with modest computational capabilities \cite{b1}, Scalpel3 introduces a modular validator and reassembly framework, reassembly heuristics informed by locality and file structure, integrated support for learned models through the ONNX Runtime~\cite{onnxruntime}, and support for fragmented file recovery through a massively threaded architecture.

Scalpel3 avoids exhaustive search by combining an architecture designed for high performance with
constraints and heuristics derived from the structure of each supported file type. Many file types,
including executable binaries, images, and compressed archives, contain internal patterns and
structural constraints that guide reconstruction and validation. Strategies that exploit this
information prune the search space and make fragmented recovery practical.

\subsection{Contributions}
\label{sec:contributions}

Scalpel3 has three primary goals: to provide a general-purpose, extensible framework in which both
contiguous and fragmented carving strategies can be implemented, tested, and deployed; to lower
the barrier between published carving techniques and tools that practitioners can use; and to recover both contiguous and fragmented data as quickly and accurately as possible on modern hardware. Performance
and practicality support these goals: speculative fragmented reassembly is tractable only on a fast
backend, and real investigations require checkpointing, operator control, and interoperability with
other tools. Throughout the paper, high performance means short wall-clock time to as many
validated, recovered files as possible and scalable reduction of the required time on platforms with more
resources (specifically, physical cores and RAM).

The contributions of this paper are:

\begin{itemize}
    \item An extensible, massively parallel carving architecture in which researchers implement single-threaded validation and reassembly logic for individual file types, while the backend supplies thread pools, I/O, synchronization, and checkpointing. Researchers do not create or manage threads (Section~\ref{sec:arch-overview}).
    \item Public application programming interfaces (APIs) for block validators, file validators, per-block and per-candidate state, and custom reassembly strategies, with ONNX support for learned models within these components. New file types are added through these interfaces without modifying the backend (Section~\ref{sec:adding-new-file-types}).
    \item A toolchain for reproducible experiments and hybrid workflows: a generator for synthetic disk images with controlled fragmentation, blockmap construction and maintenance utilities, and a FUSE filesystem that exposes only the uncovered portion of an image to other tools (Section~\ref{sec:scalpel3-toolchain}).
    \item Case studies showing how the structural signals available in a file format, such as checksums, chunk structure, decodable streams, and internal metadata, translate into validation and reassembly strategies (Section~\ref{sec:file-type-specific-carving}).
    \item An extensive performance evaluation on a corpus of more than 80{,}000 files, covering contiguous recovery with zero and random padding and fragmented recovery with gaps and out-of-order block placement, while measuring performance across increasing thread counts (Section~\ref{sec:performance}).
    \item An open source release of the complete system and toolchain under the GPLv3.
\end{itemize}

To our knowledge, Scalpel3 is the first publicly available carving framework that combines general-purpose extensibility, recovery of both contiguous and fragmented files, and massively parallel execution.

The remainder of the paper is organized as follows. Section~\ref{sec:related-work} reviews the history of file carving and recent advances, motivating the need for a general high-performance framework. Section~\ref{sec:arch-overview} outlines Scalpel3’s architecture, core data structures, and recovery phases. Section~\ref{sec:adding-new-file-types} introduces the public APIs for extending Scalpel3 with new file types. Section~\ref{sec:scalpel3-toolchain} details the Scalpel3 toolchain, including dataset generation, blockmap maintenance utilities, and blockmap-based hybrid workflows. Section~\ref{sec:file-type-specific-carving} presents case studies for representative file types. Section~\ref{sec:performance} contains our performance study. Section~\ref{sec:limitations} discusses limitations. Section~\ref{sec:future-work} discusses directions for continued research, and Section~\ref{sec:conclusions} concludes the paper.

\section{Related Work}
\label{sec:related-work}

Over the past twenty-five years, file carving research has developed along three complementary paths: (1) signature-based recovery of largely contiguous files, (2) fragmented file recovery, and (3) statistical and machine learning (ML) methods for fragment classification. Each addresses a different part of the recovery pipeline, and all three shaped Scalpel3's design.

\subsection{Header--footer Carving}

\emph{Header--footer} file carvers identify candidate files by scanning for known header and footer signatures and extracting the intervening bytes. Early tools such as Foremost~\cite{Kornblum2002Foremost} performed header--footer carving using configuration-defined signatures with minimal structural validation. Scalpel v1 (2005), based on Foremost 0.69, introduced substantial performance optimizations while retaining a largely contiguous carving model~\cite{b1}. While simple and effective for contiguous recovery, such approaches are prone to false positives when signatures are not paired with strong format checks. Scalpel development was rapid in this period: experimental versions added multithreading and GPU-accelerated searches~\cite{MarzialeRichardRoussev2007GPU}, and Scalpel v2 shipped GPU-accelerated header/footer search. OpenForensics later applied GPU pattern matching to signature carving~\cite{Bayne2018OpenForensics}. \emph{In-place} carving was explored in ScalpelFS and the independent CarvFS effort to reduce wasted output I/O and storage when many candidates are produced~\cite{RichardRoussevMarziale2007InPlace,CarvFS2009}. In parallel, PhotoRec emerged as a widely used open source, signature-based carver with broad file type coverage and consistency checks tailored to individual formats~\cite{photorec}. IPED, an open source forensic platform, includes a signature-based carver~\cite{IPED}. For recovery of unknown files, however, these tools' standard carving paths generally assume contiguity and offer limited support for fragmentation (PhotoRec's optional brute-force mode, developed for a DFRWS carving challenge, handles some fragmented JPEGs).

Scalpel3 builds on this lineage while addressing the limitations of prior tools. Unlike purely signature-based carvers, Scalpel3 introduces an extensible framework for both contiguous and fragmented file reassembly. Instead of parallelizing only select phases, Scalpel3 parallelizes everything: header/footer detection, block validation, file reassembly, file validation, and I/O.

\subsection{Fragmented File Carving} 

\paragraph{General fragmented carving}
File carving research has increasingly focused on strategies for reconstructing fragmented files~\cite{Poisel2011Multimedia}. One of the earliest approaches was \emph{bifragment gap carving} (BGC), introduced by Garfinkel (2007)~\cite{b4}. BGC targets the bifragment case in which a recognizable header and footer are present, but the contiguous region between them fails validation, suggesting the file is split by an intervening gap of unrelated sectors. The method enumerates candidate gap sizes (and consistent split points), reassembles trial candidates, and validates each until a consistent reconstruction is found. Garfinkel motivates BGC using bifragmented JPEGs from the DFRWS 2006 Carving Challenge, where one or more sectors of unrelated data appear between fragments. BGC addresses a restricted fragmentation model: it assumes a file can be represented as two fragments separated by an unknown gap, and it is most effective for formats with strong internal validation. Out-of-order fragmentation receives less attention but is common: in a study of 220 in-use Windows laptops, nearly half of fragmented files were fragmented out of order~\cite{VanDerMeer2021Fragmentation}. Huijsmans et al. (2026) analyze that dataset and present a proof-of-concept implementation for out-of-order JPEG carving, together with a tunable test image generator~\cite{Huijsmans2026OOO}. Their prototype uses known cluster membership in place of a deployable file validator, leaving practical validation and efficient out-of-order carving as open problems.

\paragraph{Format-specific validators}
General fragmented carving exposes a core challenge: fragmentation creates too many candidate reconstructions. As the number of fragments grows, the number of plausible ways to order fragments, choose join points, and account for gaps increases rapidly, making exhaustive search impractical. For this reason, modern approaches rely on format-aware validators that enforce format constraints (e.g., required structures, internal offsets and lengths, checksums such as cyclic redundancy checks (CRCs), coding constraints, and cross-field consistency) to eliminate incorrect candidates early and cheaply, substantially reducing the effective search space.

\paragraph{Image recovery (JPEG/PNG)}

Much of the work on format-specific validation for fragmented recovery has focused on image formats, particularly JPEG. Early approaches emphasized \emph{content continuity} rather than byte signatures: building on context-based document fragment reassembly~\cite{ShanmugasundaramMemon2003}, Memon and Pal (2006) modeled extracted fragments as nodes in a graph with seam-compatibility weights and posed reassembly as a maximum-weight path problem, using heuristics to manage the NP-hard search space~\cite{PalMemon2006}. Pal, Sencar, and Memon (2008) extended this line by adding sequential hypothesis testing for fragmentation point detection, combined with the Parallel Unique Path (PUP) attachment heuristic from Memon and Pal's 2006 work, greedily attaching fragments using seam metrics and reporting strong recovery on the DFRWS challenge datasets~\cite{PalSencarMemon2008SHT,DFRWS2006ChallengeRepo,DFRWS2007ChallengeRepo}. This line of work was subsequently commercialized as SmartCarving in Digital Assembly's Adroit Photo Forensics~\cite{DHSAdroit2014}.

Sencar and Memon (2009) addressed missing headers, constructing pseudo headers with encoding parameters borrowed from related images on the same medium~\cite{SencarMemon2009}, and Uzun and Sencar (2015) extended this to carving \emph{orphaned} JPEG fragments~\cite{UzunSencar2015}. Later work explored alternative ways to localize fragmentation points and order fragments, including coherence-of-Euclidean-distance (CED) block similarity~\cite{TANG2016S108} and bitstream-level invalidation signals during entropy decoding~\cite{VANDERMEER2024301687}. Durmus, Korus, and Memon (2019) push the orphaned setting further, combining photo-response non-uniformity (PRNU)-based camera attribution with interval-graph assembly to reduce compatibility testing, enabling reconstruction under severe interleaving when a camera PRNU fingerprint is available~\cite{Durmus2019JPEG}.

JPGcarve automates single-fragment and multifragment JPEG recovery~\cite{DeBockDeSmet2016}, and JpgScraper integrates JPEG data classification into a carver with released source code~\cite{UzunSencar2020}. Beyond JPEG, Hilgert \emph{et al.} carve PNGs syntactically, using chunk structure and CRCs, in phases ordered from easy to expensive cases; their prototype reassembles multifragment and out-of-order PNGs, and both the carver and Woodblock, their dataset generation framework, were released as open source~\cite{HILGERT2019S22}.

\paragraph{Beyond images}
Fragmented carving has also extended beyond images. Bin-Carver (2012) presents an early proof-of-concept for carving fragmented ELF executables without filesystem metadata~\cite{b3}. It uses signature-based searches to locate main metadata structures and constraints from the program and section header tables to bound plausible file regions. To reconnect noncontiguous blocks, Bin-Carver uses function call evidence by identifying direct \texttt{call} patterns and exploiting relationships between callers and callees, with recognizable function prologues and procedure linkage table cues, to infer block placement without full disassembly. In practice, linkage is strongest in code-heavy regions; other sections are recovered more opportunistically when they expose strong structural cues. OOXML documents have received syntactical and hybrid classifier-based reconstruction methods, with no released implementation~\cite{Boiko2023OOXML,Boiko2025OOXML}. Fragmented video recovery has its own lineage: Defraser and DC3Carver recover and repair fragmented video streams (Defraser's source remains downloadable though dormant since 2016)~\cite{CaseyZoun2014}, and research systems have addressed DVR fragments, content-based ordering of out-of-order fragments, and missing H.264 parameter sets~\cite{ParkLee2014DVR,Fang2020VideoCarving,AltinisikSencar2021H264}, with no released implementations. Fragmented MP3 recovery has used frame structure with VBR bitrate statistics~\cite{Sajja2010MP3} and similarity between the MDCT coefficients of neighboring frames~\cite{Taneja2012MP3}.

These works span diverse strategies for fragmented recovery, from bifragment gap models to executable linking guided by function call evidence, and several produced implementations. Poisel's multimedia file carver implemented the SmartCarving phases with NCD collation, PUP reassembly, and parallel per-file reconstruction paths~\cite{Poisel2011Multimedia,Poisel2013Classification}; ReviveIt, an earlier SmartCarving proof of concept from the DFRWS 2006 and 2007 challenges, was released as open source~\cite{ReviveIt2007}; Resurrection applied a three-phase carving model in a prototype JPEG carver~\cite{Lambertz2014Resurrection}; Excavator generated format validators from declarative Derric descriptions and performed bifragment gap carving on a 1~TB image~\cite{VanDenBos2012Excavator}; Cohen (2007) formalized carving and applied structure-based reconstruction to PDF and ZIP~\cite{Cohen2007Advanced}; DECA implements decision-theoretic carving with data detectors for JPEG and PNG~\cite{GladyshevJames2017,Odogwu2020PNG}; and LAYR combines metadata-based reconstruction with carving in a modular framework~\cite{Schneider2020LAYR}.

None provides a usable general tool: ReviveIt and Derric are available as source, but Excavator and the multimedia file carver are not; the remaining systems are single-format or limited prototypes. None combines publicly obtainable source code, format-neutral extension interfaces, content-based recovery of previously unknown fragmented files, and parallel execution across the major carving stages at demonstrated scale. Scalpel3 provides that combination in a high-performance, open source framework, integrating format-aware validators and reassembly functions with massively threaded scheduling, checkpointing, and I/O so that new recovery strategies can be used in practical investigations.

\subsection{File Fragment Classification}
\label{fragment-classification}

Alongside fragmented reassembly, a large body of work has studied \emph{file fragment classification}, which aims to assign an unknown block or fragment to a likely file type. Fragment classification is a useful supporting capability for fragmented carving because it can guide candidate search: for example, blocks with higher confidence for a target file type can be tried earlier when extending candidates of that type.

Content-based file type identification began with whole files: McDaniel and Heydari fingerprinted complete files using byte-frequency analysis and cross-correlation~\cite{mcdaniel2003content}. Early fragment classification applied lightweight statistical signatures to blocks and clusters, including centroid-based byte-distribution models~\cite{karresand2006oscar}, statistical disk-cluster and fragment classification~\cite{Veenman2007ClusterClassification, Calhoun2008Fragments}, and compression-based similarity measures such as Normalized Compression Distance~\cite{axelsson2010normalised}. Later work adopted richer feature representations (for example, ``bag-of-bytes'' models with discriminative classifiers~\cite{fitzgerald2012using}) and incorporated limited context to better handle compound or embedded content~\cite{sportiello2012context}. Classical supervised pipelines, often based on support vector machines (SVMs), provided strong baselines in this era (e.g., Sceadan)~\cite{beebe2013sceadan}, and hierarchical schemes were explored to improve fine-grained discrimination~\cite{bhatt2020hierarchy}.

From the late 2010s onward, researchers increasingly applied deep learning to fragment classification. Convolutional neural networks (CNNs) were applied to learn features directly from raw bytes, replacing hand-designed features~\cite{wang2018ftcnn}, and a parallel line of work converted byte blocks into grayscale images to leverage vision inductive biases~\cite{chen2018filefragment}. FiFTy substantially increased scale by releasing FFT-75 (75 file types) alongside a compact CNN classifier~\cite{mittal2020fifty}. Subsequent work refined efficiency and representations, including separable convolutions~\cite{saaim2022light}, learned byte embeddings with nearest neighbor classification~\cite{haque2022byte2vec}, and hybrid recurrent/convolutional architectures~\cite{ByteRCNN2023}. Transformer architectures continue this line, applying joint self-attention across sector boundaries~\cite{Wang2024Fusion} and Swin-based classification on FFT-75~\cite{Guzhov2025Transformer}; ByteSCAN slices fragments into fixed-size byte segments so that one model classifies fragments of any size, with released code~\cite{ByteSCAN2026}.

Despite these advances, fragment classification is usually evaluated in isolation on shuffled labeled blocks. The integrations that exist are format-specific: Qiu et al. feed SVM classification into PUP reassembly for multimedia files~\cite{Qiu2014Carving}, RX\_myKarve combines ELM classification with genetic-algorithm reassembly for JPEGs~\cite{AliMohamad2021}, JPEG-Restorer pairs syntactic JPEG carving with thumbnail-affinity checks~\cite{Birmingham2017}, and JpgScraper classifies JPEG data inside its carver~\cite{UzunSencar2020}. Scalpel3 generalizes this integration through pluggable classifiers whose block confidence scores can complement format-specific validation and guide reassembly.

\Needspace{0.9\textheight}
\section{Scalpel3: Architectural Overview}
\label{sec:arch-overview}

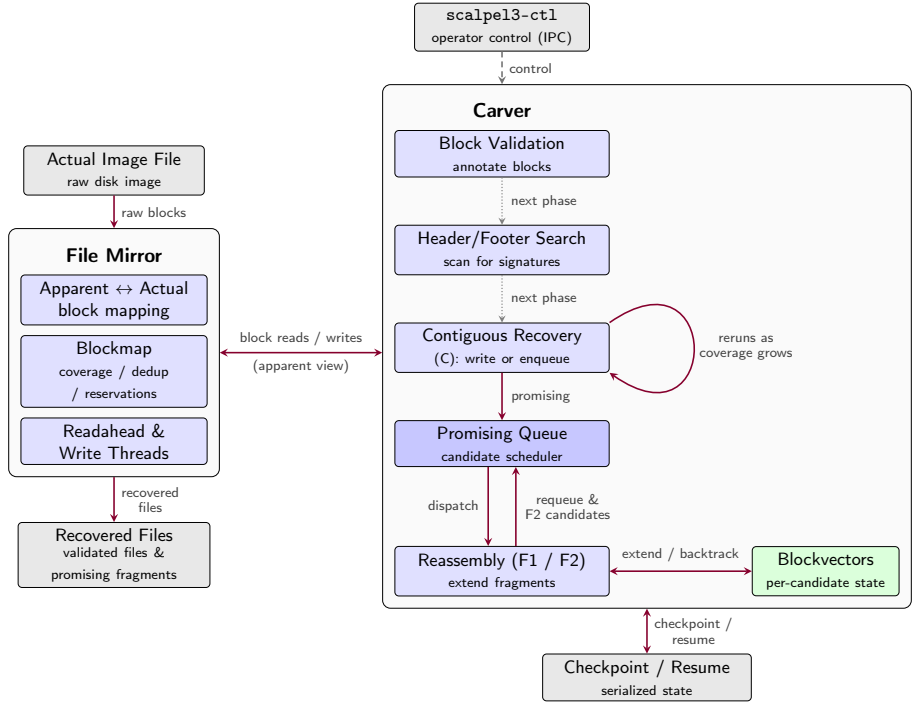
\begin{figure}[ht]
  \centering
  \resizebox{\textwidth}{!}{
  \begin{tikzpicture}[
    font=\sffamily\small,
    comp/.style={draw, rounded corners=2pt, fill=blue!12, align=center,
                 minimum height=0.6cm, inner sep=3pt, font=\sffamily\small},
    io/.style={draw, rounded corners=2pt, fill=gray!18, align=center,
               minimum height=0.65cm, inner sep=3pt, font=\sffamily\small},
    ttl/.style={font=\sffamily\bfseries},
    arr/.style={->, thick, >=stealth, color=purple!70!black},
    darr/.style={<->, thick, >=stealth, color=purple!70!black},
    ctrl/.style={->, thick, >=stealth, dash pattern=on 3pt off 2pt, color=black!55},
    seq/.style={->, semithick, >=stealth, densely dotted, color=black!50},
    fl/.style={font=\sffamily\scriptsize, text=black!75, align=center},
  ]
    \node[comp, text width=3.1cm] (aa) at (0,0.4) {Apparent $\leftrightarrow$ Actual\\block mapping};
    \node[comp, text width=3.1cm, below=5pt of aa] (bm) {Blockmap\\{\scriptsize coverage / dedup / reservations}};
    \node[comp, text width=3.1cm, below=5pt of bm] (mio) {Readahead \& Write Threads};
    \node[ttl, above=3pt of aa] (ft) {File Mirror};
    \begin{scope}[on background layer]
      \node[draw, rounded corners=4pt, fill=gray!4, fit=(ft)(aa)(bm)(mio), inner sep=6pt] (fm) {};
    \end{scope}

    \node[comp, text width=3.6cm] (bval) at (6.9,3.0) {Block Validation\\{\scriptsize annotate blocks}};
    \node[comp, text width=3.6cm, below=24pt of bval] (hf) {Header/Footer Search\\{\scriptsize scan for signatures}};
    \node[comp, text width=3.6cm, below=24pt of hf] (cr) {Contiguous Recovery\\{\scriptsize (C): write or enqueue}};
    \node[comp, fill=blue!22, text width=3.6cm, below=24pt of cr] (pq) {Promising Queue\\{\scriptsize candidate scheduler}};
    \node[comp, text width=3.6cm, below=40pt of pq] (re) {Reassembly (F1 / F2)\\{\scriptsize extend fragments}};
    \node[comp, fill=green!14, text width=2.4cm, right=72pt of re] (bvec) {Blockvectors\\{\scriptsize per-candidate state}};
    \node[ttl, above=3pt of bval] (ct) {Carver};
    \begin{scope}[on background layer]
      \node[draw, rounded corners=4pt, fill=gray!4, fit=(ct)(bval)(hf)(cr)(pq)(re)(bvec), inner sep=6pt] (carver) {};
    \end{scope}

    \node[io, text width=3.0cm, above=0.8cm of ft] (aif) {Actual Image File\\{\scriptsize raw disk image}};
    \node[io, text width=3.2cm, below=0.8cm of fm.south] (rec) {Recovered Files\\{\scriptsize validated files \&\\promising fragments}};
    \node[io, text width=2.9cm, above=0.8cm of ct] (ctl) {\texttt{scalpel3-ctl}\\{\scriptsize operator control (IPC)}};
    \node[io, text width=3.5cm, below=0.8cm of carver.south] (ckpt) {Checkpoint / Resume\\{\scriptsize serialized state}};

    \draw[arr] (aif.south) -- (aif.south|-fm.north) node[midway,right,fl]{raw blocks};
    \draw[arr] (fm.south-|rec.north) -- (rec.north) node[midway,right,fl]{recovered\\files};
    \draw[ctrl] (ctl.south) -- (ctl.south|-carver.north) node[midway,right,fl]{control};
    \draw[darr] (carver.south-|ckpt.north) -- (ckpt.north) node[midway,right,fl]{checkpoint /\\resume};
    \draw[darr] (fm.east) -- (fm.east-|carver.west)
      node[midway,above,fl]{block reads / writes}
      node[midway,below,fl]{(apparent view)};

    \draw[seq] (bval) -- (hf) node[midway,right=1pt,fl]{next phase};
    \draw[seq] (hf) -- (cr) node[midway,right=1pt,fl]{next phase};
    \draw[arr] (cr) -- (pq) node[midway,right=1pt,fl]{promising};
    \draw[arr] ([xshift=-7pt]pq.south) -- ([xshift=-7pt]re.north) node[midway,left=1pt,fl]{dispatch};
    \draw[arr] ([xshift=7pt]re.north) -- ([xshift=7pt]pq.south) node[midway,right=1pt,fl]{requeue \&\\F2 candidates};
    \draw[darr] (re) -- (bvec) node[midway,above=1pt,fl]{extend / backtrack};

    \draw[arr] (cr.north east) to[out=35,in=-35,looseness=7] (cr.south east);
    \node[fl, anchor=west] at ([xshift=42pt]cr.east) {reruns as\\coverage grows};
  \end{tikzpicture}
  }
  \caption{High-level architecture of Scalpel3. The file mirror maps between actual and apparent block numbers, maintains the blockmap, and mediates I/O. The carver annotates blocks, discovers file boundaries, validates contiguous candidates, and schedules partially validated candidates for fragmented reassembly. Each validated recovery updates coverage, simplifying the apparent image for subsequent passes. The \texttt{scalpel3-ctl} utility provides operator control, and restart checkpoints preserve recovery state.}
  \label{fig:scalpel-arch}
\end{figure}

Scalpel3 is a multithreaded file carving framework designed for high-performance recovery of fragmented files from disk images and other storage media. Its architecture consists of several coordinated components that identify, reconstruct, and validate fragments using format-specific knowledge at both the block and file levels. The core components include the carver, file mirror, block and file validators, and supporting data structures such as blockvectors, blockmaps, and the promising queue. Figure~\ref{fig:scalpel-arch} summarizes the architecture, and Table~\ref{tab:terminology} defines the principal terms used throughout the paper.

\begin{table}[H]
\centering
\footnotesize
\caption{Principal architectural terms.}
\label{tab:terminology}
\begin{tabular}{@{}l p{8.6cm}@{}}
\toprule
Term & Meaning \\
\midrule
Block & Fixed-size unit of recovery; the granularity at which data is tracked and reassembled \\
Actual image & The raw disk image; an actual block number is a physical position in it \\
Apparent image & The logical view with covered blocks removed; positions renumber as coverage grows. Uncovered non-exemplar duplicates remain visible to contiguous recovery \\
Covered block & A block currently excluded from the apparent image, typically after validated recovery or an explicit blockmap update \\
Duplicate, exemplar & Identical blocks are coalesced for fragmented recovery; the exemplar is the single representative, while contiguous recovery retains uncovered physical copies \\
Reservation & Count of in-progress reassembly candidates currently holding a block; soft coverage that discourages, but does not forbid, reuse \\
Blockmap & Per-block store of coverage, duplication, and reservation state; the primary copy provides the covered view, and a shadow copy accumulates updates that are merged at synchronization points, including between successful contiguous passes and at periodic checkpoints \\
Blockvector & A candidate's ordered sequence of chosen blocks, with the apparent-to-actual mapping and per-slot alternatives \\
Promising candidate & A partially validated candidate awaiting reassembly in the promising queue \\
Promising queue & Priority queue that schedules promising candidates by accumulated service time \\
File mirror & The I/O layer that maintains the actual and apparent views and mediates all block reads and writes \\
MoDiCo & Scalpel3's custom learned block classifier, which assigns initial block type predictions \\
C, F1, F2 & The recovery phases: contiguous recovery, reassembly of promising candidates, and creation of speculative candidates from unused headers and footers \\
\texttt{validates\_to} & Inclusive byte offset of the last byte known to be format-consistent for a reassembly candidate; reported by file validators \\
Periodic checkpoint & Merges accumulated shadow blockmap updates into the primary blockmap, so later recovery passes see new coverage \\
Progress checkpoint & Publishes in-progress reassembly work for examination by operators and visualization tools \\
Restart checkpoint & Serializes restartable state to disk so carving can resume with little lost work \\
\bottomrule
\end{tabular}
\end{table}

\Needspace{4\baselineskip}
The carver coordinates both contiguous and fragmented recovery, operating on a virtualized view of the disk image maintained by the file mirror. The subsections that follow describe each component; Section~\ref{sec:running-example} traces them end to end on a small example image.

\subsection{The Carver}

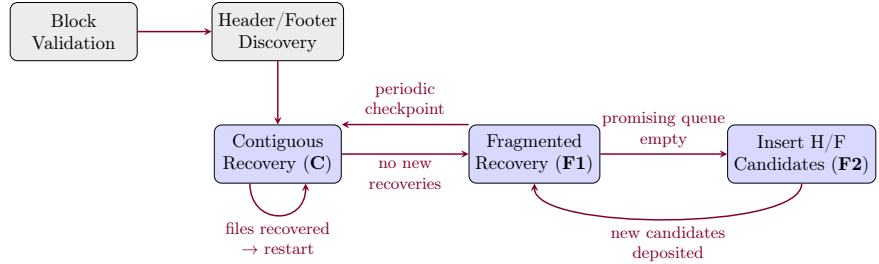
\begin{figure}[ht]
  \centering
  \scalebox{0.7}{
  \begin{tikzpicture}[
    init/.style={draw, rounded corners=4pt, fill=gray!15, minimum width=2.4cm,
                 minimum height=1.1cm, align=center, font=\normalsize},
    phase/.style={draw, rounded corners=4pt, fill=blue!15, minimum width=2.4cm,
                  minimum height=1.1cm, align=center, font=\normalsize},
    arr/.style={->, thick, >=stealth, color=purple!70!black},
    lbl/.style={font=\small, align=center},
    node distance=0.8cm
  ]
    \node[init] (BV) {Block\\Validation};
    \node[init, right=1.4cm of BV] (HF) {Header/Footer\\Discovery};

    \node[phase, below=1.2cm of HF] (C) {Contiguous\\Recovery (\textbf{C})};
    \node[phase, right=2.4cm of C] (F1) {Fragmented\\Recovery (\textbf{F1})};
    \node[phase, right=2.4cm of F1] (F2) {Insert H/F\\Candidates (\textbf{F2})};

    \draw[arr] (BV) -- (HF);
    \draw[arr] (HF) -- (C);

    \draw[arr] ([xshift=-15pt]C.south) .. controls ++(0,-0.8) and ++(0,-0.8) ..
    node[below, lbl, align=center] {files recovered\\$\to$ restart} ([xshift=15pt]C.south);

    \draw[arr] (C.east) -- node[below, lbl] {no new\\recoveries} (F1.west);

    \draw[arr] (F1.north west) -- node[above, lbl, pos=0.5] {periodic\\checkpoint} (C.north east);

    \draw[arr] (F1.east) -- node[above, lbl] {promising queue\\empty} (F2.west);

    \draw[arr] (F2.south) .. controls ++(0,-0.9) and ++(0,-0.9) ..
      node[below, lbl] {new candidates\\deposited} (F1.south);
    \end{tikzpicture}
    }
  \caption{Scalpel3 processing phases. Block validation and
    header/footer discovery (gray) run once as initialization.
    Contiguous recovery (C) then restarts each time files are recovered,
    since covering blocks may close gaps. Fragmented recovery (F1)
    periodically pauses for a C retry, then resumes. When the promising queue
    is empty during reassembly, F2 deposits previously unused headers/footers as
    single-block candidates for F1.}
      \label{fig:scalpel-phases}
\end{figure}

The carver coordinates the search, validation, and reassembly threads that reconstruct files from a disk image. First, block validators identify and annotate candidate blocks for each file type. Next, header and footer identification locates candidate file boundaries, using either regular expressions or custom discovery functions. After this one-time initialization, recovery proceeds through three iterative phases: C, F1, and F2. Figure~\ref{fig:scalpel-phases} summarizes the coordination of these recovery phases.

\begin{itemize}
    \item C (Contiguous recovery): The carver attempts to recover files that are logically contiguous. Recovery begins from a discovered header or custom boundary, constructs a candidate extent using footer evidence, format metadata, or configured limits, and validates the candidate. Since the file mirror hides already covered blocks, logical contiguity can be revealed even if the file is fragmented. Uncovered duplicate positions remain visible during contiguous recovery so that deduplication does not alter the physical layout being tested. Successfully validated files are written out, and their blocks marked as covered. If validation fails but a partial structure is detected, the candidate is placed in the promising queue for further analysis in F1. Phase C is fast and repeats because covering each newly validated file can reveal additional logically contiguous candidates.

    \item F1 (Fragmented reassembly based on promising fragments): This phase handles partially validated candidates that exhibit structural evidence but do not yet form complete files. Extending these candidates is generally more expensive than C, but their existing structural evidence makes the search more constrained than in F2. The carver attempts to extend candidates until they validate or are discarded. At checkpoints that merge blockmap updates, C recovery is restarted when new files have validated, allowing newly exposed logical contiguity to be exploited before fragmented reassembly resumes.

    \item F2 (Fragmented reassembly based on remaining headers/footers): When the promising queue has no candidates available for dispatch, F2 creates new candidates from remaining unprocessed headers or footers. Active F1 candidates may still be running; the new candidates enter the same queue and keep otherwise idle reassembly threads supplied with work. Because these candidates begin with little structural evidence, their search space is large and may require substantial backtracking. F2 therefore gives priority to more constrained promising candidates before adding speculative work.
\end{itemize}

\vspace{4pt}

The phases prioritize easy recoveries first, remove the blocks of validated files from the search space, and escalate progressively to more expensive strategies. The phases are not strictly linear: the system repeatedly transitions between them as block coverage evolves. During F1, reassembly threads pause at periodic checkpoints (Section~\ref{sec:checkpointing}): the primary blockmap is updated to reflect newly covered blocks, and active candidates are returned to the promising queue with priorities adjusted for their accumulated service time before Phase~C is retried. During fragmented recovery, candidates are scheduled by accumulated service time so work is distributed across the promising queue.

These phases are driven by dedicated thread pools and shared candidate queues. Header/footer discovery populates the database used by C to construct candidates; candidates that validate only partially enter the promising queue before dispatch to a reassembly thread. The carver manages these transitions using priority queues and several dedicated thread pools:

\begin{itemize}
    \item Header/footer search: At startup, this pool performs lightweight boundary scans based on signatures, magic bytes, or other distinguishing patterns. The results populate the header/footer database, which later supplies entry points for contiguous recovery and, when needed, F2 candidate creation. The scan uses portable single instruction, multiple data (SIMD) operations, with NEON on ARM processors and AVX2/AVX-512 on Intel/AMD processors where available; in our measurements, vectorized scanning is faster than a scalar implementation by up to 2.2$\times$ on Apple Silicon and 4.7$\times$ with AVX-512. Simple literal signatures and alternations in regular expressions that can be reduced to literal patterns use the SIMD path. Patterns requiring custom discovery functions or general regular expression matching are handled by conventional search threads.

    \item File and block validation: Invokes block validators (early in execution) and file validators (throughout all phases) to assess structural correctness of file candidates and promising fragments.

    \item Reassembly: Invokes the core reconstruction logic that extends fragments, calling validators, reassembly functions, and performing backtracking. This thread pool also monitors a kill queue to terminate candidates marked for abandonment by the operator or deemed unnecessary or duplicative by the carver.
\end{itemize}

Finally, the carver allows for checkpointing and operator control. Specifically, state can be flushed to disk and threads halted using global synchronization primitives, so that carving can resume from a restart checkpoint. Through the \texttt{scalpel3-ctl} utility, operators can issue commands to checkpoint, query status, or abandon specific candidates if necessary. This feature provides human-in-the-loop control of long-running jobs.

\subsection{A Running Example}
\label{sec:running-example}

\begin{figure}[t]
  \centering
  \resizebox{\textwidth}{!}{%
  \begin{tikzpicture}[
    font=\sffamily,
    blk/.style={draw, minimum width=0.95cm, minimum height=0.9cm, inner sep=1pt,
                font=\sffamily\small, align=center},
    pone/.style={blk, fill=cyan!22},
    ptwo/.style={blk, fill=green!18},
    jp/.style={blk, fill=orange!22},
    dupb/.style={blk, fill=violet!18},
    oth/.style={blk, fill=black!12},
    num/.style={font=\sffamily\footnotesize, text=black!75},
    sub/.style={font=\sffamily\footnotesize, text=black!75, align=center},
  ]
    \node[oth]  (b0) {};
    \node[pone, right=0pt of b0]  (b1) {\textbf{H}};
    \node[pone, right=0pt of b1]  (b2) {};
    \node[jp,   right=0pt of b2]  (b3) {\textbf{H}};
    \node[jp,   right=0pt of b3]  (b4) {\textbf{F}};
    \node[pone, right=0pt of b4]  (b5) {\textbf{F}};
    \node[ptwo, right=0pt of b5]  (b6) {\textbf{H}};
    \node[ptwo, right=0pt of b6]  (b7) {};
    \node[dupb, right=0pt of b7]  (b8) {};
    \node[oth,  right=0pt of b8]  (b9) {};
    \node[ptwo, right=0pt of b9]  (b10) {};
    \node[ptwo, right=0pt of b10] (b11) {\textbf{F}};
    \node[dupb, right=0pt of b11] (b12) {};
    \node[oth,  right=0pt of b12] (b13) {};

    \foreach \i in {0,...,13} {
      \node[num, below=2pt of b\i.south] {\i};
    }

    \begin{scope}[shift={($(b0.south west)+(0.0,-1.3)$)}]
      \node[pone, minimum width=0.5cm, minimum height=0.5cm, anchor=west] (l1) {};
      \node[sub, right=3pt of l1, anchor=west] (l1t) {PNG-1};
      \node[jp, minimum width=0.5cm, minimum height=0.5cm, right=0.7cm of l1t.east, anchor=west] (l2) {};
      \node[sub, right=3pt of l2, anchor=west] (l2t) {JPEG};
      \node[ptwo, minimum width=0.5cm, minimum height=0.5cm, right=0.7cm of l2t.east, anchor=west] (l3) {};
      \node[sub, right=3pt of l3, anchor=west] (l3t) {PNG-2};
      \node[dupb, minimum width=0.5cm, minimum height=0.5cm, right=0.7cm of l3t.east, anchor=west] (l4) {};
      \node[sub, right=3pt of l4, anchor=west] (l4t) {duplicate filler};
      \node[oth, minimum width=0.5cm, minimum height=0.5cm, right=0.7cm of l4t.east, anchor=west] (l5) {};
      \node[sub, right=3pt of l5, anchor=west] {unrelated data};
    \end{scope}
  \end{tikzpicture}%
  }
  \caption{Layout of the running example image: fourteen 16~KiB blocks. \textbf{H} and \textbf{F} mark discovered header and footer positions. PNG-1 is fragmented around the contiguous JPEG; PNG-2 is fragmented around a filler block that recurs at block 12 and an unrelated block of compressed data.}
  \label{fig:running-example}
\end{figure}
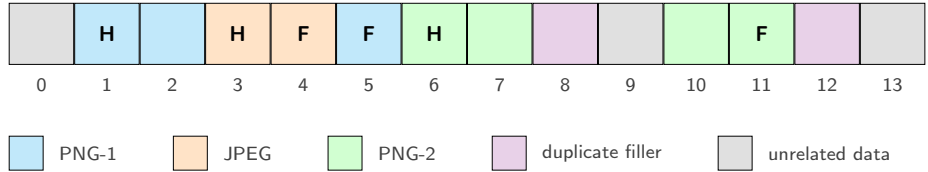

The example image has fourteen 16~KiB blocks (Figure~\ref{fig:running-example}). Blocks 3--4 hold a small contiguous JPEG. A three-block PNG, PNG-1, is split around it: blocks 1--2 and block 5. A second, four-block PNG, PNG-2, occupies blocks 6--7 and 10--11, separated by a filler block (8) whose content recurs at block 12 and by an unrelated block of compressed data (9). Blocks 0 and 13 are unrelated. Throughout, we track each candidate's blockvector, written as the sequence of actual block numbers it holds, and the coverage (C), duplicate (D), exemplar (E), and reservation (T) entries of the blockmap (Section~\ref{sec:blockmaps}); Table~\ref{tab:running-example} summarizes the state after each step.

\begin{enumerate}
    \item \emph{Initialization.} Blockmap construction finds that blocks 8 and 12 are identical: 8 becomes the exemplar and 12 a non-exemplar duplicate that fragmented recovery will never see. Block validation begins with MoDiCo, the learned block classifier (Section~\ref{sec:machine-learning}), which produces content class confidence scores for each exemplar block; duplicate blocks inherit the scores assigned to their exemplar. Blocks belonging to the PNGs and JPEG receive high scores for their respective types, while the filler and unrelated blocks receive low scores for each type. None is excluded because MoDiCo prioritizes blocks but never invalidates them. The block validators run next; each receives MoDiCo's confidence and may retain or change it while recording additional block state. The PNG block validator records each block's CRC32, which becomes part of the index that reassembly consults later (Section~\ref{sec:file-type-specific-carving}). Header/footer discovery then populates the database: PNG signatures at blocks 1 and 6, IEND footers at 5 and 11, and the JPEG pair at 3 and 4.

    \item \emph{First contiguous pass.} Phase~C pairs headers with footers and validates the apparent spans. The JPEG (blocks 3--4) validates and is written; its blocks are marked covered. The PNG-1 candidate spans blocks 1--5 and fails when validation reaches JPEG data: the validator's per-block CRC checkpoints identify the end of block 2 as the last consistent boundary, so the candidate is truncated to the promising fragment $\langle 1,2 \rangle$ and enqueued. The PNG-2 candidate (blocks 6--11) is truncated the same way to $\langle 6,7 \rangle$ and enqueued.

\end{enumerate}

The next four subsections each end with an \emph{Example} paragraph that advances the trace through the mechanism just described.

\begin{table}[t]
\centering
\small
\caption{State changes in the running example. Blockvectors list actual block numbers; C, D, and E are the blockmap coverage, duplicate, and exemplar flags, and T is the reservation counter (Section~\ref{sec:blockmaps}).}
\label{tab:running-example}
\begin{tabular}{@{}l p{2.9cm} p{2.6cm} p{4.7cm}@{}}
\toprule
Step & Event & Blockvectors & Blockmap and queue effects \\
\midrule
1 & Dedup, block validation, discovery & none & D, E on 8; D on 12; MoDiCo confidence scores recorded; per-block CRCs stored as PNG block state \\
2 & First C pass & $\langle 1,2 \rangle$, $\langle 6,7 \rangle$ created & JPEG written, 3--4 covered; both candidates enqueued; T set on 1, 2, 6, 7 \\
3 & Coverage merged, C reruns & $\langle 1,2 \rangle$ removed as stale & PNG-1 validates contiguously (actual 5 is now apparent 3); 1, 2, 5 covered; T cleared on 1, 2 \\
4 & F1 extension & $\langle 6,7 \rangle \rightarrow \langle 6,7,10 \rangle$ & 8 and 9 excluded by CRC arithmetic; 10 verified and reserved \\
5 & Completion & $\langle 6,7,10,11 \rangle$ written & 6, 7, 10, 11 covered; reservations released; queue empty; F2 unused \\
\bottomrule
\end{tabular}
\end{table}

\subsection{The Promising Queue}

The promising queue is a priority queue that schedules partially validated file candidates for fragmented reassembly. Candidates are enqueued once they exhibit structural promise (e.g., valid headers or partial validation) but do not yet validate as complete files. Each candidate carries a scheduling priority that is initialized when the candidate is created and serialized across restart checkpoints. When a checkpoint returns an active candidate to the queue, the service time consumed since dispatch is subtracted from its priority, so candidates that have received less service rise toward the front. A priority assigned to each file type can additionally bias which types are serviced first; it is disabled by default.

This policy helps inexpensive candidates validate quickly while expensive candidates are progressively deprioritized rather than monopolizing the pool, and it surfaces early partial results across many files rather than complete reconstructions of a few. Reassembly threads dispatch the candidate that has received the least service, extend it using format-specific logic, and invoke validators: successful reconstructions are written and their blocks marked covered, shrinking the remaining search space, while unsuccessful candidates are requeued at reduced priority or abandoned if their universally unique identifier (UUID) appears in the kill queue. Periodic maintenance enforces operator kill requests and removes stale entries.

\paragraph{Example} The truncated fragments $\langle 1,2 \rangle$ and $\langle 6,7 \rangle$ enter the queue at step 2 of the trace (Table~\ref{tab:running-example}), each carrying its initial priority based on accumulated service time. Queue maintenance removes $\langle 1,2 \rangle$ at step 3, when coverage makes it stale, so F1 services only $\langle 6,7 \rangle$. Had the queue emptied while unused headers or footers remained, F2 would have created speculative candidates from those boundaries and placed them in the same queue (F2 changes how and when candidates are created, not how they are subsequently reconstructed). In this trace, F2 never runs because every discovered header and footer has been consumed when $\langle 6,7 \rangle$ completes and the queue becomes empty at step 5.

\subsection{Blockvectors}

In Scalpel3, each reassembly candidate maintains its own blockvector, an ordered sequence of block references representing the candidate's current logical layout. A blockvector serves as the working context for validating, extending, and writing files, and has two primary roles:

\begin{itemize}
    \item Reconstruction workspace: It records logical block order and associated metadata. Candidate bytes are accessed through the mapped image when possible or materialized into a contiguous buffer when required.

    \item Apparent-to-actual block mapping: In collaboration with the file mirror, it maps apparent block numbers (positions in the file mirror's current logical view after covered blocks are masked) to actual block numbers (block positions in the original image). During fragmented recovery, duplicate candidate choices are separately coalesced through their exemplars.
\end{itemize}

Each slot in the blockvector contains the following:
\begin{itemize}
    \item An apparent block number, identifying the selected block in the file mirror's current apparent image. The slot's blockvector index determines its logical position in the reconstructed file.
    \item An actual block number, denoting its position in the original image.
    \item A validity flag, indicating whether the associated block of data has been read and is usable.
    \item In addition, each slot maintains a choices structure that tracks the candidate blocks already tried and those that remain to be tried for that blockvector index.
\end{itemize}

Reassembly strategies access the apparent-to-actual mapping through the blockvector API. They can reason about logical file order, such as filling slots from left to right, while preferring blocks physically near previous selections. This supports locality-sensitive placement and backtracking without repeatedly consulting global metadata.

Scalpel3's default left-to-right (LR) reassembly strategy shows how blockvectors are used during execution. When extending a candidate, a block still available for the next blockvector slot is selected and appended to the candidate. If the resulting configuration fails validation or does not meaningfully improve the candidate, that block is recorded as tried for that slot, preventing it from being retried in the same position. In addition, the reassembly logic maintains a per-candidate queue of the best block choices for the current slot, which is consulted during backtracking to revisit the most promising alternatives. When an appended block validates through its end, LR attempts to extend the candidate across consecutive blocks in the apparent image, using galloping to test increasingly long runs. This quickly places regions that are logically contiguous after covered blocks are removed. The combination lets Scalpel3 explore many candidate configurations while avoiding redundant reads and minimizing validation overhead.

Although each blockvector belongs to one candidate, the same block may appear in multiple blockvectors across file types or candidate fragments. A block remains available until it belongs to a fully validated file and is marked covered. This permits parallel exploration of overlapping reconstructions.

\paragraph{Example} At step 4, a reassembly thread extends $\langle 6,7 \rangle$. The incomplete final \texttt{IDAT} chunk in block 7 declares its length, which locates the chunk's stored CRC. Using the per-block CRCs recorded during block validation (Section~\ref{sec:file-type-specific-carving}), the PNG strategy filters the candidate pool arithmetically for uncovered, PNG-plausible blocks that could complete the chunk consistently with that target. The filler exemplar 8 and the unrelated block 9 are excluded without a validator call; block 10 matches, is confirmed by the usual inflate and filter checks, and fills the next slot, giving $\langle 6,7,10 \rangle$. At step 5, block 11 supplies the remaining structure and the \texttt{IEND} footer, and $\langle 6,7,10,11 \rangle$ validates fully and is written. Formats without such a deterministic constraint extend by trial instead: a block that makes incremental validation regress is recorded as tried in the slot's choices structure, the preceding blockvector is restored, and the next candidate block is tried. This is the backtracking used by the left-to-right strategy described above; the trace never needs it.

\subsection{The File Mirror}
\label{sec:file-mirror}

The file mirror mediates access to the raw disk image. It maintains both the actual image, which preserves the physical block layout, and a dynamically updated apparent image with covered blocks removed. Uncovered duplicate positions remain visible to contiguous recovery, while fragmented recovery coalesces duplicate choices through their exemplars. This gives threads a consistent logical view while preserving references to actual disk blocks. The file mirror provides three principal services:

\begin{itemize}
    \item Apparent vs. actual address translation: The file mirror maintains a mapping between actual block numbers (physical location on the disk image) and apparent block numbers (logical view used during carving). This mapping, accessed through the mirror API, allows Scalpel3 components to navigate the shifting apparent space as covered blocks are removed while retaining references to their actual positions in the evidence image.

    \item Duplicate and covered block management: Each block in the image is tracked through metadata describing coverage and duplication status. The file mirror exposes functions to query whether a block is covered, a duplicate, or an exemplar copy. Covered blocks are excluded from recovery to avoid redundant or conflicting reconstructions. During fragmented recovery, duplicate blocks are coalesced through their exemplar to avoid redundant validation effort; contiguous recovery retains uncovered physical copies in their original positions.

    \item I/O abstraction and optimization: The file mirror mediates reads and writes through a uniform interface. A sequential readahead thread reduces latency during scans, while a vector write pool handles recovered output asynchronously. Candidate data may be accessed directly through memory mappings or materialized into a contiguous buffer when required.
\end{itemize}

Scalpel v1 and v2 were engineered to conserve memory, which was scarce on the machines of their era, while still carving quickly. Scalpel3 uses available memory more aggressively to improve performance. It accesses the evidence image through one read-only memory mapping that is shared by all threads. The operating system retains as many recently accessed image pages as available memory permits and can discard those pages when memory is needed. The image therefore does not need to fit in physical memory, while additional RAM allows more of the image to remain cached. Scalpel3 separately allocates metadata proportional to the number of image blocks and registered file types; for example, the block confidence table uses one byte for each combination of an image block and an active file type. Each worker also maintains bounded working buffers, so memory use increases with thread count. A configuration option reduces aggressive allocation on systems with limited memory.

\begin{figure}[htbp]
    \centering
    \includegraphics[scale=0.447]{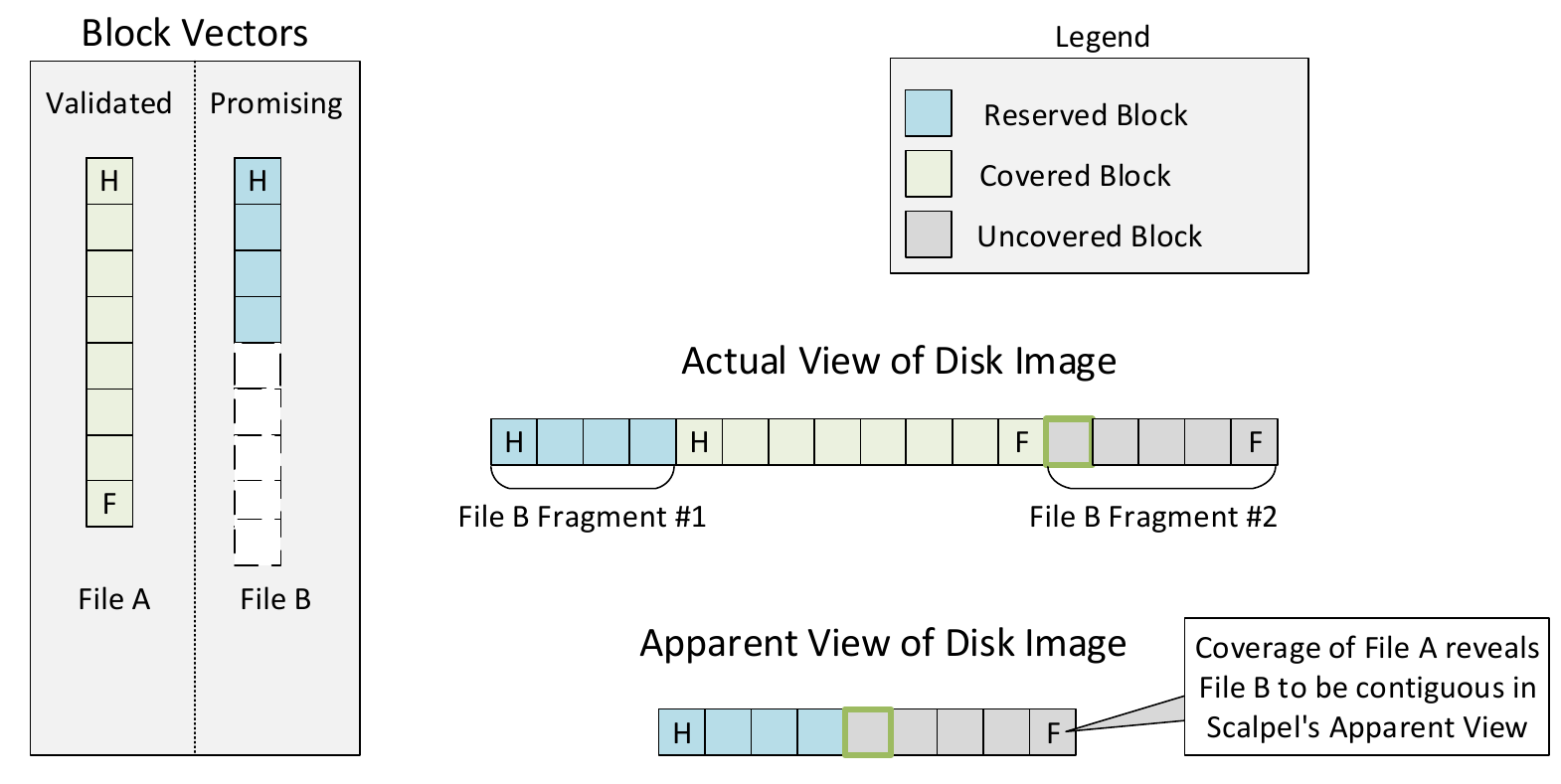}
    \caption{Interaction between blockvectors, the actual disk layout, and the apparent logical view. When File A is validated and its blocks are covered, the remaining fragments of File B align contiguously in the apparent image, enabling a trivial contiguous recovery effort.}
    \label{fig:blockvector}
\end{figure}

Figure~\ref{fig:blockvector} shows how the actual and apparent images interact. Each file candidate maintains a blockvector that tracks its selected blocks. The apparent image omits validated file data while preserving mappings to the corresponding blocks in the actual image. When \texttt{File A} validates and its blocks are covered, \texttt{File B} becomes contiguous in the apparent image, enabling fast recovery under mild fragmentation.

The file mirror is tightly integrated with blockvectors, which represent the working state of file candidates. Blockvectors rely on the file mirror to translate apparent and actual block numbers, fetch block data, and update coverage during reconstruction.

\paragraph{Example} Step 3 of the trace is the file mirror at work. Covering the JPEG's blocks 3 and 4 removes them from the apparent image, so blocks 1, 2, and 5 become consecutive in the apparent view (actual block 5 is now apparent block 3), and the retried header at block 1 reaches its footer without interruption: PNG-1 validates completely, a fragmented file recovered by contiguous logic alone, exactly the File~A/File~B interaction of Figure~\ref{fig:blockvector}. This coverage feedback provides a cheap path for mild fragmentation: the second contiguous pass recovers PNG-1 without invoking fragmented reassembly.

\subsection{Blockmaps}
\label{sec:blockmaps}

The \textit{Blockmap} is a bitmap-like data structure maintained by the file mirror to track block coverage, duplication, and reservation status throughout reassembly. Its purpose is threefold: (1) to implement the coverage system controlled by the file mirror, (2) to eliminate redundant computation by identifying duplicate blocks and exemplar relationships, and (3) to support Scalpel3's block reservation system.

\subsubsection{Deduplication and Reservations}
\label{reservations}

Modern disk images frequently contain redundant data, such as filler blocks (e.g., $0x00$ or $0xFF$), repeated content from backups or cloned partitions, or incidental duplicates of valid file data. Without deduplication, Scalpel3 would repeatedly try identical blocks in reassembly candidates, wasting computation on equivalent failures. Conversely, naively discarding duplicates risks breaking valid contiguous fragments.

To address this, Scalpel3 implements a blockwise deduplication model that differentiates between exemplars (the single representative for a set of duplicated blocks), covered duplicates, and uncovered duplicate blocks. Each duplicate remains visible at its actual position during contiguous recovery, while fragmented reassembly considers only the exemplar for the group. The blockmap includes four bits and two integer values per block.

\begin{itemize}
    \item Covered (C): marks blocks currently excluded from recovery, typically because they belong to a fully validated file or were explicitly covered through the blockmap toolchain.
    \item Duplicate (D): flags blocks that are identical to one or more blocks in the image.
    \item Exemplar (E): indicates whether a block is the representative instance among a group of duplicates.
    \item Zero (Z): indicates whether a block is an all-zero block (e.g., zero padding or sparse regions).
    \item Reference Count (R): for exemplars, the number of uncovered blocks in the duplicate group, including the exemplar itself; for non-exemplars, the index of their exemplar. Uncovered nonduplicate blocks carry $R = 1$, representing a single available instance; covering the block sets $R = 0$.
    \item Reservations (T): a counter on each exemplar recording how many in-progress candidates currently hold the block or one of its duplicates.
\end{itemize}

Table~\ref{tab:blockmap} summarizes how combinations of C, D, and E are interpreted by Scalpel3's backend in combination with R. Exemplar blocks represent duplicate groups and carry shared state; reference counts keep an exemplar available to fragmented recovery until every occurrence in its duplicate group is covered.

\begin{table}[h!]
\centering
\caption{Semantics of C/D/E flags in the blockmap: how coverage (C), duplication (D), and exemplar (E) status determine each block $j$’s state.}
\label{tab:blockmap}
\begin{tabular}{@{}lll p{9cm}@{}}
\toprule
\textbf{C} & \textbf{D} & \textbf{E} & \textbf{Interpretation}                                                                                                                                                                    \\ \midrule
0          & 0          & 0          & Uncovered, non-duplicate block. $R[j] = 1$, a single uncovered instance ($R$ is unused when deduplication is disabled).                                                                                                              \\
0          & 1          & 0          & Uncovered duplicate, non-exemplar. Seen only by contiguous recovery. $R[j]$ stores the exemplar's block index; the corresponding reference count is $R[R[j]]$.                                          \\
0          & 1          & 1          & Uncovered exemplar block, representing one or more uncovered duplicates. Seen by all recovery phases. If $R[j]=0$, the block and all of its duplicates are effectively covered. \\
1          & 0          & 0          & Covered, non-duplicate block. No deduplication metadata.                                                                                                                                   \\
1          & 1          & 0          & Covered, non-exemplar duplicate. No longer visible.                                                                                                                                        \\
1          & 1          & 1          & Covered exemplar.                                                                                                                  \\ \bottomrule
\end{tabular}
\end{table}

Reservations implement a form of soft coverage that reduces contention among reassembly threads. When a block enters a candidate’s blockvector, its exemplar’s reservation counter $T$ is incremented; when it leaves, $T$ is decremented. Through reservations, reassembly strategy developers can choose to prefer blocks with smaller values of  $T$, while others remain eligible but are deprioritized. This logic is used by Scalpel3's default reassembly strategy, since it reduces fruitless validator calls on blockvectors containing blocks that are likely to be validated and covered soon by other candidates. However, this mechanism is optional and can be applied in a strategy-dependent manner to best support custom reassembly policies.

\subsubsection{Blockmap Updates}

\begin{figure}[ht]
    \centering
    \includegraphics[trim=35 200 24 224, clip, width=\linewidth]{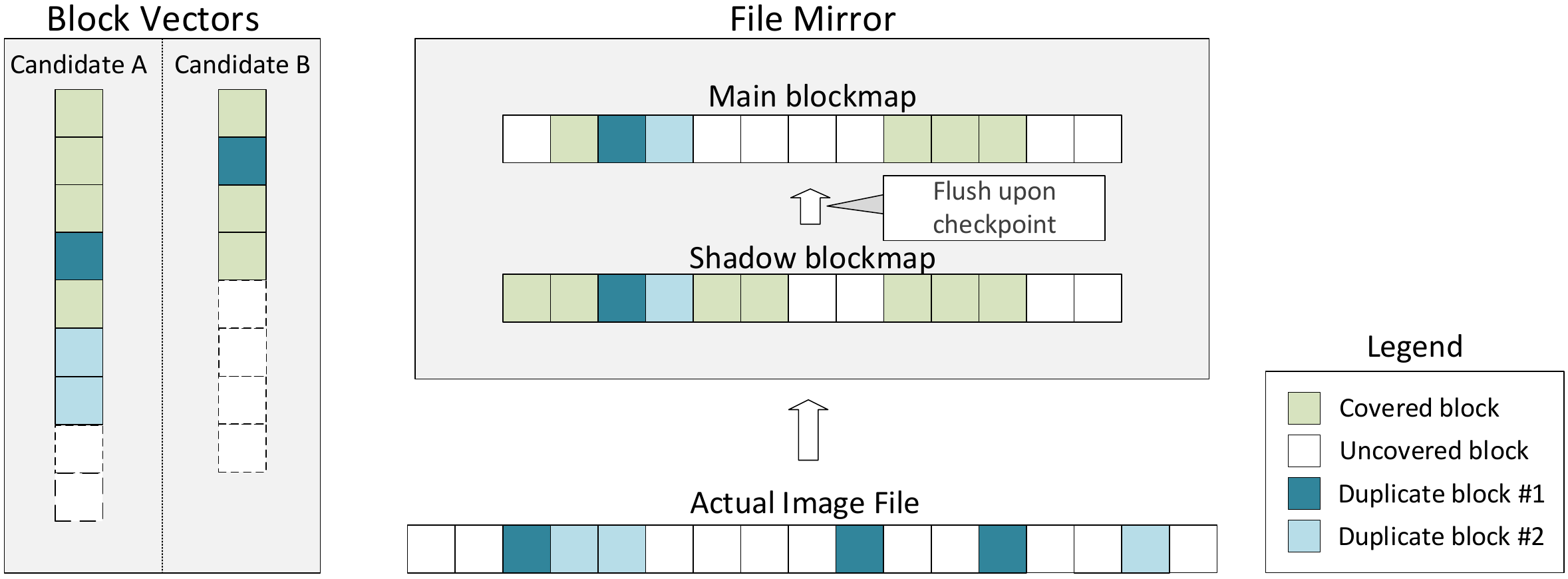}
    \caption{Scalpel3's primary and shadow blockmaps during fragmented reassembly. Candidate blockvectors at left contain selections from the actual image shown below; repeated colors identify blocks with duplicate content. Coverage updates accumulate in the shadow blockmap and are merged into the primary blockmap at a periodic checkpoint, producing the refreshed view used by recovery threads.}
    \label{fig:blockmaps}
\end{figure}

To reduce contention, Scalpel3 maintains both a primary blockmap and a shadow blockmap. During reassembly, coverage updates accumulate in the shadow map, and block selection consults it to avoid blocks already claimed in the current recovery epoch. At a periodic checkpoint, reassembly threads become quiescent, the shadow map becomes the new primary map, and the apparent-to-actual mappings are rebuilt. This design avoids a global blockmap lock while still allowing fine-grained updates to be batched efficiently. Figure~\ref{fig:blockmaps} shows promising candidates under reconstruction alongside the primary and shadow blockmaps. The actual image appears below with duplicate and covered blocks visible.

\paragraph{Example} The blockmap carries the trace's bookkeeping. At step 1, deduplication marks block 8 as the exemplar for its duplicate at block 12 (D and E set on 8, D on 12), so fragmented recovery never considers block 12. At step 2 the truncated candidates set $T = 1$ on blocks 1, 2, 6, and 7; removing stale $\langle 1,2 \rangle$ at step 3 releases blocks 1 and 2, and step 4 reserves block 10. Coverage accumulates in the shadow blockmap as files validate. After step 2, the ordinary blockmap swap between passes merges those updates before C runs again, shrinking the apparent view and enabling PNG-1's contiguous recovery at step 3. At step 5 the written file's blocks are covered and its reservations released; every block of both PNGs and the JPEG ends covered, and the filler exemplar survives uncovered with $R = 2$.

\subsection{Validators}

Block validators assess whether individual blocks are consistent with a given file type. When MoDiCo is enabled, block validators run after its classification pass (Section~\ref{sec:machine-learning}) and may retain or change its per-block confidence scores. Because rejecting even one true file block can prevent successful reassembly, block validators minimize false negatives; the same block may therefore be accepted for multiple file types. Beyond type membership, a block validator can record finer-grained labels and other per-block metadata through the block-state API (Section~\ref{block-state}); the per-block CRCs that drive the running example's reassembly step (Section~\ref{sec:running-example}) are recorded this way. File validators evaluate the structural integrity of assembled fragments at the file level: in contiguous recovery they test entire logically contiguous candidates, in left-to-right (LR) reassembly they incrementally check the fragment as each block is appended, and custom reassembly strategies invoke them at points of their choosing. Both kinds are developer-extensible through the APIs of Section~\ref{sec:adding-new-file-types}.

\subsubsection{Machine Learning}
\label{sec:machine-learning}

For some file types, reliable block or file validation is difficult to implement using manually designed structural checks alone. In such settings, developers may incorporate machine learning (ML) approaches, including the fragment classification methods discussed in Section~\ref{fragment-classification}. Scalpel3 integrates the ONNX Runtime~\cite{onnxruntime} into its validator and custom reassembly interfaces, so that learned models can be invoked during carving. The primary built-in consumer is MoDiCo, a learned block content classifier that runs at the start of the block validation phase, before any block validator. MoDiCo predicts per-class confidence scores for each exemplar block; duplicate blocks inherit the confidence scores assigned to their exemplar. MoDiCo confidence scores influence the order in which candidate blocks are considered during reassembly; they do not remove blocks from the search space. Each block validator receives the current confidence, may retain or change it, and can record additional state specific to the format. MoDiCo can be disabled per run, and inference runs on the CPU or on available accelerators. Learned inference supplies evidence that complements structural validation and guides reassembly decisions. Extending learned validation and reassembly to additional file types, and evaluating these techniques at scale, remain active research directions; MoDiCo's development is described in~\cite{Hildebrand2026}, and the interpretability of its predictions is examined in~\cite{McCain2026MoDiCo}.

\subsection{Checkpointing}
\label{sec:checkpointing}

Scalpel3 uses three kinds of checkpoints. Periodic checkpoints update the blockmap: updates accumulated in the shadow blockmap are merged into the primary blockmap, so later recovery passes see newly covered blocks. Progress checkpoints publish in-progress reassembly work for examination by operators and visualization tools. Restart checkpoints are scheduled every 5m by default and serialize restartable state to disk, including candidate queues, blockvectors, blockmap metadata, carving progress, and format-specific validator and carve state.   Restart checkpoints save practitioners time by avoiding redundant computation, safeguard progress against unexpected interruptions such as power outages, and let investigators pause and resume long jobs deliberately with little lost work.  A checkpoint and exit request writes a restart checkpoint when restartable state exists and then shuts down cleanly; before restartable state exists, Scalpel3 stops without writing a checkpoint. Only restart checkpoints concern recovery after termination; the other two support ongoing runs.

\subsection{Human in the Loop Carving}

\label{human-in-the-loop}

When processing large disk images with many fragmented files, even modern systems with hundreds of cores and well-designed validators cannot guarantee timely completion or alignment with case-specific priorities. To address this limitation, Scalpel3 incorporates an interactive, human-in-the-loop control facility, allowing an operator to monitor and adjust carving operations in real time.

This interface allows analysts to tailor carving based on domain knowledge or investigative priorities. Operators can inspect reassembly jobs in progress, monitor block coverage, review statistics, and terminate jobs that are unlikely to yield useful results. Because reassembly jobs are tracked by UUID, cancellation is reliable and consistent throughout the system. The interprocess communication (IPC) facility that supports this command-line interface is also used by Scalpel3's graphical visualization facility, which can display block coverage, inspect partially recovered files, and interactively manage carving progress.

\subsection{Scalpel3 Output}

Scalpel3 outputs results incrementally. Fully validated files are written as soon as validation succeeds, while partially recovered fragments can optionally be emitted for inspection. Outputs are organized under status directories (\texttt{VALIDATED/}, \texttt{PROMISING/}, and \texttt{INPROGRESS/}), with optional subdirectories by file type. Validated and promising filenames include the primary and clone UUIDs and a SHA-256 hash of the contents; in-progress filenames use the UUID pair and retain a stable pathname for monitoring. When blockvector output is enabled, Scalpel3 also records the list of actual blocks used, enabling deeper auditability and post hoc analysis.

\section{Adding New File Types}
\label{sec:adding-new-file-types}

\subsection{Validator API}

To add support for contiguous or fragmented recovery of a new file type, developers implement a small collection of single-threaded functions. These include a required \textit{file validator} and, for many formats, a \textit{block validator}. Block validators determine whether individual blocks may belong to the type, while file validators assess the validity of fragments at the file level. Block validators are optional because some formats are better handled through header/footer discovery and file-level validation alone, while others require block-level filtering to make fragmented recovery practical.

\subsubsection{Block Validators}

Scalpel3 records an integer confidence score from 0 to 100 for each block and active file type. A block validator receives the current score, inspects the contents of the block, and may preserve or update the score. The score is an eligibility and ordering signal, not a calibrated probability. Confidence values are interpreted as follows:

\begin{itemize}
    \item Invalid (0): the block cannot belong to the file type.
    \item Low confidence (1): the block may belong to the file type, but no stronger evidence is available.
    \item Graded confidence (2--99): increasing confidence, such as that supplied by a classifier.
    \item Maximum confidence (100): a definitive positive assignment, or the permissive default when no applicable block evidence is available.

\end{itemize}

To minimize false negatives, which can prevent entire files from validating, block validators are expected to err on the side of inclusiveness. Ambiguous blocks are marked as potentially valid and are later disambiguated by file validators or reassembly logic. Block validators may also record auxiliary state, such as offsets and structural or section information. Scalpel3 stores this state in an associated hash structure, which becomes read-only once block validation completes. The block state API is discussed below.

\subsubsection{File Validators}

File validators examine the structure and content of fragments and produce three outputs:

\begin{itemize}
    \item \textit{promising}: true if the fragment shows partial validity, false if completely invalid.
    \item \textit{validates\_to}: the inclusive byte offset of the last byte known to be consistent with the file format.
    \item \textit{validates}: true if the fragment fully validates as a complete file, false otherwise.
\end{itemize}

These outputs directly guide reassembly. In contiguous recovery (C), a file validator tests suspected contiguous files. When a candidate does not fully validate but is marked \textit{promising}, Scalpel3 uses \textit{validates\_to} to retain a shorter prefix and places it in the promising queue; a candidate that is neither valid nor promising is discarded. In fragmented reassembly (F1/F2), the file validator may be invoked incrementally as new blocks are appended, ensuring that reconstruction proceeds only along format-consistent paths.

\subsection{State API}
\label{sec:blockstateapi}

Scalpel3 provides state APIs that associate custom metadata with individual blocks and candidates under reconstruction. They preserve format-specific context across validation, reassembly, and checkpointing.

\subsubsection{Block State}

\label{block-state}

The block state API associates metadata with individual blocks. This is useful for encoding information such as block type, subheader presence, or other structural hints discovered during validation. Block state may be updated throughout block validation. Once that phase completes, it becomes immutable, ensuring consistent interpretation of the block across reassembly attempts. Internally, state objects are retrieved through a block hash key that abstracts over duplicates and exemplars, ensuring consistent access to state information throughout reassembly. The running example (Section~\ref{sec:running-example}) depends on this contract: block validation records each block's CRC32 as block state, and the F1 extension step consults those values long after the phase that produced them has completed.

To support extensibility, developers provide a small set of type-specific functions for managing state, including serialization/deserialization (for checkpointing), cloning, and freeing, along with optional sizing and debugging helpers. These hooks allow Scalpel3 to manage block state safely while giving developers flexibility to encode domain-specific information.

\subsubsection{Carve State}

The carve state API manages metadata at the level of a reassembly candidate rather than a single block. Unlike block state, carve state is mutable throughout the carving process, allowing validators and reassembly logic to track progress, accumulate metadata, or record the presence of format-specific structures. Validators and reassembly threads access carve state through a unique carve key.

Carve state uses the same management hooks as block state, so candidate metadata can be serialized and restored across checkpoints.

\subsection{Custom Reassembly Functions}

Custom reassembly functions give developers direct control over fragmented recovery. They use Scalpel3's core data structures and backend APIs to implement format-specific block selection, validation, backtracking, and checkpoint handling.

A custom reassembly function controls a candidate's lifecycle within a backend reassembly thread, from initialization through final output:

\begin{itemize}
    \item Initialization: setting up a candidate's metadata and blockvector.
    \item Block selection and placement: identifying plausible blocks for extension based on validator results, duplication status, locality, or domain-specific heuristics.
    \item Validation and commitment: integrating new blocks into the candidate, updating state, and invoking file validators to confirm structural consistency.
    \item Candidate backtracking: restoring prior state and trying alternative block choices after validation fails.
    \item Checkpoint handling: returning control promptly when a checkpoint is requested and preserving the candidate state needed for serialization and resumption.
\end{itemize}

Reassembly functions can also use optional facilities such as \textit{galloping mode}, in which multiple sequential blocks are tentatively placed and the number tested grows exponentially after each successful validation. If validation fails, the candidate is trimmed back to the last verified state and resumes conservative block-by-block exploration. This exploits the long contiguous runs commonly found within fragmented files and is used by Scalpel3's default reassembly strategy. Reassembly functions may also use the reservation system discussed in Section~\ref{reservations} to prefer blocks not held by active candidates. The default strategy uses reservations to avoid redundant validator calls on blocks likely to belong to other candidates.

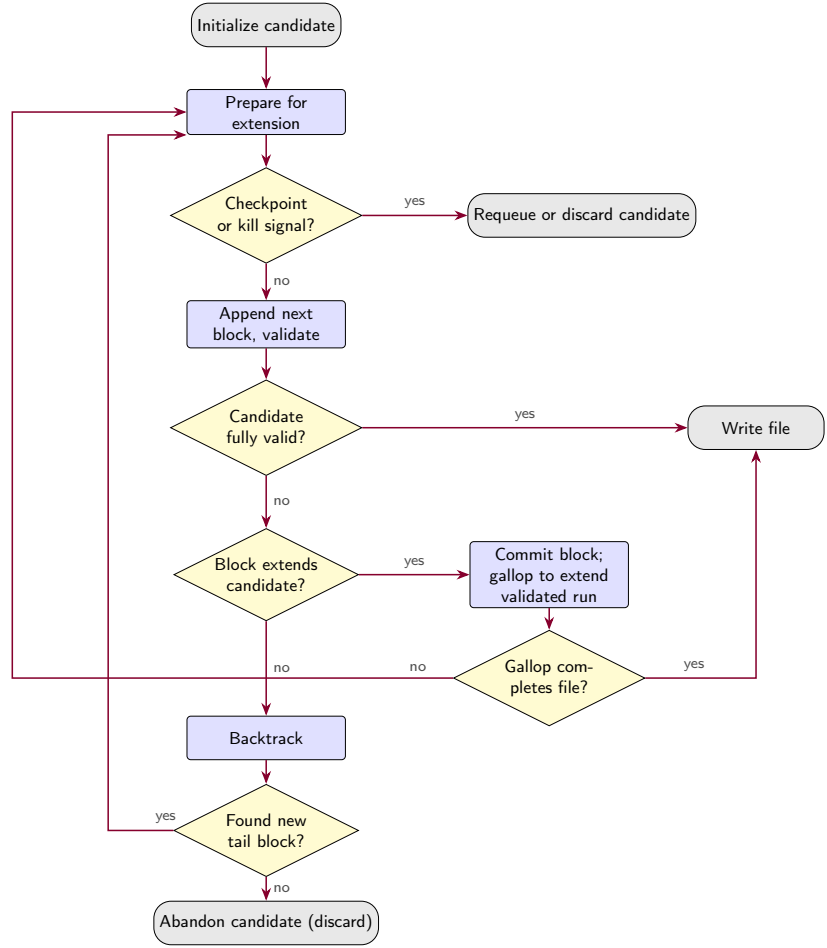
\begin{figure}[t]
  \centering
  \scalebox{0.72}{
  \begin{tikzpicture}[
    font=\sffamily\small,
    term/.style={draw, rounded corners=9pt, fill=gray!18, minimum height=0.8cm,
                 minimum width=2.5cm, align=center, inner sep=3pt},
    proc/.style={draw, rounded corners=2pt, fill=blue!12, minimum height=0.8cm,
                 text width=2.7cm, align=center, inner sep=3pt},
    dec/.style={draw, diamond, aspect=2.0, fill=yellow!22, align=center,
                inner sep=1pt, text width=1.95cm},
    arr/.style={-{Stealth[length=2.4mm]}, thick, color=purple!70!black},
    lbl/.style={font=\sffamily\footnotesize, text=black!70},
  ]
    \node[term] (t0) at (0,0)      {Initialize candidate};
    \node[proc] (p1) at (0,-1.6)   {Prepare for extension};
    \node[dec]  (d1) at (0,-3.5)   {Checkpoint or kill signal?};
    \node[proc] (p2) at (0,-5.5)   {Append next block, validate};
    \node[dec]  (d2) at (0,-7.4)   {Candidate fully valid?};
    \node[dec]  (d3) at (0,-10.1)  {Block extends candidate?};
    \node[proc] (p5) at (0,-13.1)  {Backtrack};
    \node[dec]  (d5) at (0,-14.8)  {Found new tail block?};
    \node[term] (ta) at (0,-16.5)  {Abandon candidate (discard)};

    \node[term] (tr) at (5.8,-3.5)   {Requeue or discard candidate};
    \node[term] (tw) at (9.0,-7.4)   {Write file};
    \node[proc] (p3) at (5.2,-10.1)  {Commit block; gallop to extend validated run};
    \node[dec]  (d4) at (5.2,-12.0)  {Gallop completes file?};

    \draw[arr] (t0) -- (p1);
    \draw[arr] (p1) -- (d1);
    \draw[arr] (d1) -- (p2) node[midway,right,lbl]{no};
    \draw[arr] (p2) -- (d2);
    \draw[arr] (d2) -- (d3) node[midway,right,lbl]{no};
    \draw[arr] (d3) -- (p5) node[midway,right,lbl]{no};
    \draw[arr] (p5) -- (d5);
    \draw[arr] (d5) -- (ta) node[midway,right,lbl]{no};

    \draw[arr] (d1) -- (tr) node[midway,above,lbl]{yes};
    \draw[arr] (d2) -- (tw) node[midway,above,lbl]{yes};
    \draw[arr] (d3) -- (p3) node[midway,above,lbl]{yes};
    \draw[arr] (p3) -- (d4);
    \draw[arr] (d4.east) -| (tw.south) node[pos=0.22,above,lbl]{yes};

    \draw[arr] (d4.west) -- node[pos=0.08,above,lbl]{no} ++(-8.1,0) |- (p1.west);
    \draw[arr] (d5.west) -- node[pos=0.12,above,lbl]{yes} ++(-1.2,0) |- (p1.south west);
  \end{tikzpicture}
  }
  \caption{Control flow of Scalpel3's default left-to-right (LR) reassembly. A candidate is repeatedly extended one block at a time and revalidated; \emph{galloping} accelerates extension across runs of logically contiguous blocks that validate, trimming if it overreaches. Processing of a candidate ends in one of three ways: the candidate fully validates and is written; a checkpoint returns it to the promising queue or a kill signal discards it; or backtracking is exhausted and the candidate is abandoned. A custom reassembly function may reuse this overall structure, but can also control candidate state, block selection and placement, validation cadence, checkpoint handling, backtracking, and work sharing.}
  \label{fig:LR-reassembly}
\end{figure}

The main reassembly loop coordinates these responsibilities, iterating until a candidate either validates as a complete file or is abandoned. Scalpel3’s default strategy is left-to-right (LR) reassembly, which incrementally extends a candidate by appending blocks at the tail and invoking the file validator after each extension. In the absence of a format-specific reassembly function, the LR strategy serves as the baseline. Figure~\ref{fig:LR-reassembly} summarizes the LR control flow, including validation, backtracking, and checkpoint responsiveness.

This interface lets developers add format-aware heuristics, probabilistic search strategies, or placement guided by machine learning without changing the backend.

\subsection{Registering New File Types}

At minimum, adding a new format for contiguous or fragmented recovery requires implementing a file validator and registering the format in the configuration so the carver can schedule it across the relevant search and validation phases. Block validators, custom reassembly functions, and state hooks are added when the format or the recovery strategy requires them:

\begin{itemize}
\item Block validator (optional): flags blocks that may belong to the type (recall-oriented; may attach block state).
\item File validator: checks structural validity of fragments and drives truncation and completion signals.
\item State hooks (required if using state): serialize, clone, and free functions for block and candidate state.
\item Reassembly function (optional): type-specific strategy used in F1/F2; if omitted, the left-to-right (LR) baseline is used.
\end{itemize}

Registration is declarative: a single configuration entry specifies the type name, size bounds, header and footer signatures (or custom discovery functions), the search direction, and function pointers for the components above, along with a scheduling priority and, optionally, a flag that excludes the type from fragmented recovery. Most formats set only a few of the available fields. Once the entry is compiled in, the backend schedules the new type through discovery, validation, and reassembly without further integration work. The distribution includes registration entries for all supported types.

\section{Scalpel3 Toolchain}
\label{sec:scalpel3-toolchain}

\subsection{Dataset Generation}

The Scalpel3 distribution includes \texttt{fragmentator}, a generator for synthetic disk images with controlled fragmentation. We use \texttt{fragmentator} extensively for performance studies and for developing new file type support because it makes experiments with controlled fragmentation reproducible. Alongside each image, \texttt{fragmentator} emits a key file recording the ground truth layout, allowing recovered files to be compared byte for byte with the originals.

\texttt{fragmentator} consumes a configuration file containing global image parameters and per-file placement directives. \texttt{OUTPUTFILE} names the generated image, \texttt{SEED} makes randomized placement reproducible, and \texttt{BLOCKSIZE} sets the image block size. \texttt{ZEROPADBLOCKS} and \texttt{RANDOMPADBLOCKS} pad the image to a requested total block count, whereas \texttt{ZEROBLOCKS} and \texttt{RANDOMBLOCKS} add exact numbers of standalone filler blocks. Each \texttt{FILE} directive begins a source file entry. \texttt{FRAGMENTED: TRUE} keeps any configured header blocks contiguous and separates each remaining file block into its own fragment; \texttt{MISSING} omits specified zero-based logical file blocks; \texttt{OUTOFORDER} relocates specified zero-based logical file blocks; and \texttt{GAP} inserts intervening blocks at a specified boundary. For example, \texttt{GAP: 5-7} inserts three blocks before logical block 5. Together, these directives control fragmentation, gap size, and block ordering. Figure~\ref{fig:frg} depicts a condensed configuration file.

\begin{figure}
    \centering
\small{
\begin{verbatim}
/* Global image parameters */
OUTPUTFILE: corpus.img
SEED: 42
BLOCKSIZE: 16384
ZEROPADBLOCKS: 0
RANDOMPADBLOCKS: 0
RANDOMBLOCKS: 0
ZEROBLOCKS: 0

/* Per-file directives */
FILE: file1.dat
MISSING: 1,3,5,10-20
OUTOFORDER: 2
FILE: file2.dat
FRAGMENTED: TRUE # fully fragmented after header blocks
FILE: file3.dat
FRAGMENTED: FALSE # contiguous
FILE: file4.dat
GAP: 5-7

\end{verbatim}
}
    \caption{Condensed example of a fragmentator configuration file for generating a synthetic disk image.}
    \label{fig:frg}
\end{figure}

\subsection{Blockmap Maintenance}

The Scalpel3 toolchain includes utilities for constructing and maintaining the blockmap, a per-block index that records coverage, duplication relationships, and reservation metadata:

\begin{itemize}
    \item \texttt{crblockmap}: Builds an initial blockmap for a disk image, detects all-zero blocks, and performs block-level deduplication using per-block SHA-256 hashes, with candidate duplicates confirmed by direct block comparison. The resulting coverage and deduplication metadata are stored in the blockmap for reuse across runs.
    \item \texttt{modblockmap}: Supports offline inspection and controlled edits to blockmap coverage state, including manually covering or uncovering individual blocks or ranges. It can also ingest logs from external recovery tools (e.g., PhotoRec) to mark their outputs as covered, enabling hybrid workflows in which third-party tools recover certain formats while Scalpel3 focuses on the remaining search space.
    \item \texttt{cmpblockmaps}: Compares two blockmaps to highlight changes, primarily differences in coverage, so practitioners can quantify progress (e.g., before vs.\ after a Scalpel3 run).
\end{itemize}

\subsection{Blockmapfs}

The Scalpel3 toolchain also includes \texttt{blockmapfs}, a FUSE (Filesystem in Userspace) \cite{FUSE} filesystem that overlays a disk image with its blockmap and exposes a view in which all covered blocks are removed. Downstream raw content scanners and carving tools that read the \texttt{blockmapfs} mount see only the uncovered portion of the image, avoiding redundant work on data already validated by Scalpel3 or other tools.

Combined with \texttt{modblockmap}, this enables modular, hybrid workflows:

\begin{enumerate}
    \item Run a third-party tool (\texttt{tool\_1}) to recover formats X/Y/Z.
    \item Use \texttt{modblockmap} to mark the corresponding blocks as covered (ingesting \texttt{tool\_1}'s logs if available).
    \item Run Scalpel3 on the residual space; newly validated files are also marked covered.
    \item Mount the image with \texttt{blockmapfs} to expose the remaining uncovered view.
    \item Run another tool (\texttt{tool\_2}) against this view to target formats not handled by \texttt{tool\_1} or Scalpel3.
\end{enumerate}

\section{Format-Specific Carving}
\label{sec:file-type-specific-carving}

Because file formats differ in their internal structure (e.g., headers, framing, alignment, compression, checksums), validator design must be format-specific. Some formats also require custom reassembly routines, particularly for fragmented recovery. This section describes several representative Scalpel3 validators: PNG, JPEG, and GIF, which are the fragmented image formats evaluated in Section~\ref{sec:performance}; ELF, which illustrates metadata-driven length recovery for a format without a footer; MP3, which illustrates stream recovery based on chained frame structure; and ZIP, which illustrates container validation and covers the modern Microsoft Office formats. These descriptions concentrate on behavior: the structures each strategy exploits and the evidence it uses. Many implementation details are omitted; for those, see the released code. The section closes with the path from a carving idea to a working experiment.

\subsection{PNG File Carving}
\label{sec:png-carving}

\begin{figure}[h]
    \centering
    \includegraphics[scale=0.4]{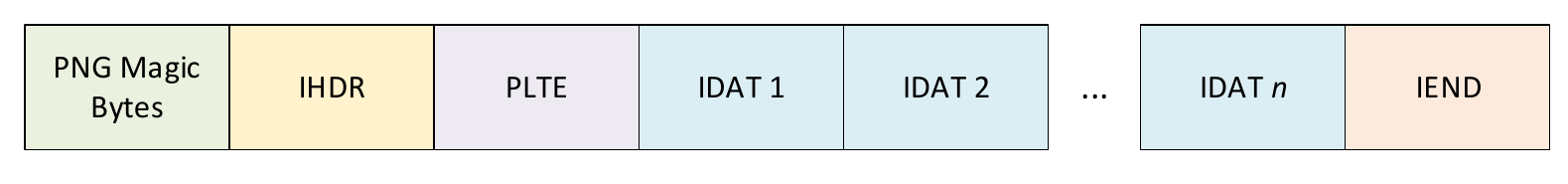}
    \caption{Simplified representation of the PNG file format.}
    \label{fig:png-format}
\end{figure}

The anatomy of a typical PNG file~\cite{PNGSpec} is depicted in Figure~\ref{fig:png-format}: a PNG begins with an 8-byte signature, followed by a sequence of length-prefixed, typed chunks, each protected by a CRC: \texttt{IHDR}, an optional \texttt{PLTE} palette, one or more \texttt{IDAT} chunks that carry the compressed image stream, and a terminating \texttt{IEND}.

The validator processes a reassembly candidate sequentially, iterating over chunks and enforcing core syntactic constraints (e.g., well-formed length and type fields, correct chunk boundaries, and matching CRCs). For \texttt{IDAT} chunks, which carry the compressed image stream, the validator performs a deeper check using a \texttt{zlib} inflate context for the validation pass. It attempts to inflate the \texttt{IDAT} payload to detect stream-level errors, verifies that the zlib parameters are valid, and checks that the decompressed scanlines use valid PNG filter types. Validation is incremental: the furthest format-consistent offset becomes \texttt{validates\_to}, and CRC checkpoints recorded at block boundaries within \texttt{IDAT} data locate the last consistent block after a chunk CRC failure, so truncation of a failing candidate is exact.

For fragmented recovery, the PNG strategy treats stored chunk CRCs as reconstruction constraints, exploiting the linearity of CRC32 over GF(2). Prior syntactical carvers use chunk CRCs to verify assembled candidates~\cite{HILGERT2019S22}; here the constraint is inverted, selecting blocks by solving for the required contribution rather than testing assemblies. Scalpel3 precomputes forward and inverse CRC shift operators for relevant byte lengths. Writing $\operatorname{shift}_t(c)$ for the contribution $c$ shifted across $t$ following bytes, a chunk with one unresolved block position determines the required contribution exactly:
\[
\operatorname{shift}_t(\mathrm{CRC}(b)) \;=\; \mathrm{CRC}_{\mathrm{stored}} \oplus C_{\mathrm{known}}
\]
where $C_{\mathrm{known}}$ combines the contributions of the already placed data. A hash table indexes the CRCs of uncovered blocks, so a single unresolved position is solved by applying the inverse shift and looking up the required block CRC. Multiple unresolved positions use structured run and anchor searches plus meet-in-the-middle decomposition; harder gaps use memory- and work-bounded exact paths and a block-by-block CRC-constrained fallback that preserves known contiguous positions. The same CRC geometry applies to ancillary and \texttt{IDAT} chunks. Every proposed placement is subsequently checked by the normal structural, inflate, and filter validation path, making acceptance of a spurious 32-bit collision substantially less likely. This gives PNG recovery a stronger signal than formats that lack per-chunk checksums: many local reassembly decisions are checked against a deterministic CRC target rather than ranked only by plausibility.

\subsection{JPEG File Carving}

\begin{figure}[h]
    \centering
    \scalebox{0.85}{
    \begin{tikzpicture}[
      seg/.style={draw, thin, minimum height=1.15cm, align=center,
                  font=\sffamily\small, inner sep=4pt},
    ]
      \node[seg, fill=green!18,  text width=1.3cm] (soi) {SOI\\{\scriptsize 0xFFD8}};
      \node[seg, fill=yellow!30, text width=1.0cm, right=0pt of soi]  (app) {APP$n$};
      \node[seg, fill=violet!15, text width=0.9cm, right=0pt of app]  (dqt) {DQT};
      \node[seg, fill=violet!15, text width=0.9cm, right=0pt of dqt]  (dht) {DHT};
      \node[seg, fill=yellow!30, text width=0.9cm, right=0pt of dht]  (sof) {SOF};
      \node[seg, fill=cyan!15,   text width=0.9cm, right=0pt of sof]  (sos) {SOS};
      \node[seg, fill=cyan!15,   text width=2.6cm, right=0pt of sos]  (scan) {Entropy-Coded\\Data};
      \node[seg, fill=orange!18, text width=1.3cm, right=0pt of scan] (eoi) {EOI\\{\scriptsize 0xFFD9}};
    \end{tikzpicture}
    }
    \caption{Simplified representation of the JPEG file format.}
    \label{fig:jpeg-format}
\end{figure}
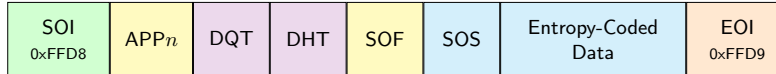

In contrast to PNG, the JPEG format provides no per-chunk checksums, which makes it the most challenging of the formats we evaluate for fragmented recovery (Section~\ref{sec:performance}). A JPEG file begins with a start-of-image marker (\texttt{0xFFD8}) and a sequence of marker segments carrying quantization and Huffman tables and a frame header, followed by a start-of-scan marker and the entropy-coded image data, and ends with an end-of-image marker (\texttt{0xFFD9}), as illustrated in Figure~\ref{fig:jpeg-format}.

The Scalpel3 JPEG validator parses this marker structure and validates the entropy-coded scan by Huffman decoding it one minimum coded unit (MCU) at a time using the file's Huffman tables, rejecting invalid codes and illegal byte stuffing. Restart markers (\texttt{RST}$n$), when present, provide resynchronization points. The last offset that decodes cleanly is recorded as \texttt{validates\_to}, and full validation requires the terminating EOI marker. The validator handles both baseline and progressive scans, cross-checks its own decoder against a \texttt{libjpeg} decoder, tolerates zero-filled runs inside and after a candidate when they do not invalidate its structure, and decodes a scaled preview to measure pixel-row continuity at block boundaries.

A dedicated JPEG reassembly strategy combines structural validation with decoded image continuity at block boundaries to rank alternative continuations. A join may be rejected even when it decodes if it produces a severe discontinuity. The strategy searches across gaps, backfills tails, and backtracks over ranked alternatives. Because entropy-coded data carries no per-block checksum and an incorrect continuation often decodes for many bytes before failing, JPEG offers weaker deterministic evidence for block ordering than checksum-constrained formats like PNG; this is the root cause of its lower recovery under out-of-order fragmentation in Section~\ref{sec:performance}.

\subsection{GIF File Carving}

\begin{figure}[h]
    \centering
    \scalebox{0.85}{
    \begin{tikzpicture}[
      seg/.style={draw, thin, minimum height=1.15cm, align=center,
                  font=\sffamily\small, inner sep=4pt},
    ]
      \node[seg, fill=green!18,  text width=1.2cm] (hdr) {Header\\{\scriptsize GIF89a}};
      \node[seg, fill=yellow!30, text width=1.45cm, right=0pt of hdr]  (lsd)  {Screen\\Descriptor};
      \node[seg, fill=violet!15, text width=1.8cm, right=0pt of lsd]  (gct)  {Global\\Color Table};
      \node[seg, fill=yellow!30, text width=1.45cm, right=0pt of gct]  (idsc) {Image\\Descriptor};
      \node[seg, fill=cyan!15,   text width=2.5cm, right=0pt of idsc] (data) {LZW Image Data\\{\scriptsize (sub-blocks)}};
      \node[seg, fill=orange!18, text width=1.1cm, right=0pt of data] (trl)  {Trailer\\{\scriptsize 0x3B}};
    \end{tikzpicture}
    }
    \caption{Simplified representation of the GIF file format.}
    \label{fig:gif-format}
\end{figure}
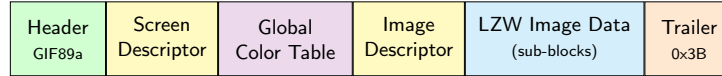

A GIF file~\cite{GIFSpec} begins with a six-byte signature (\texttt{GIF87a} or \texttt{GIF89a}) and a logical screen descriptor, optionally followed by a global color table, as shown in Figure~\ref{fig:gif-format}. The body is a sequence of one or more image blocks, each introduced by an image descriptor (\texttt{0x2C}) and an optional local color table, interleaved with extension blocks; the file ends with a single-byte trailer (\texttt{0x3B}). Within an image block, the pixel data is LZW-compressed and stored as a chain of length-prefixed \emph{sub-blocks} of one to 255 bytes each, terminated by a zero-length sub-block.

The Scalpel3 GIF validator, adapted from \texttt{giflib}'s \texttt{giffix}, validates this structure by decompressing it: a self-contained LZW decoder walks the sub-block chain, maintaining the code table and rejecting malformed codes and sub-blocks that violate the length structure. As with PNG's inflate check, a clean decode is strong evidence that the bytes belong together. Because GIF carries no per-chunk checksum, however, the validator supplements the decode with two statistical detectors that run as pixels are produced. The first tracks the mean absolute difference (MAD) between consecutive decoded scanlines: correct image data yields visually coherent rows, whereas an incorrect continuation typically produces a sharp discontinuity. The second models the LZW consumption rate (compressed bytes consumed per decoded row), which a wrong block perturbs even when its codes remain individually legal. Clean end-of-information termination is tracked as an additional integrity signal, and the furthest cleanly decoded offset is recorded as \texttt{validates\_to}.

For fragmented recovery, the length-prefixed sub-block structure supplies an additional, GIF-specific constraint. Because sub-block boundaries fall at predictable offsets within the compressed stream, the validator derives a \emph{phase}, the alignment of sub-block boundaries relative to the block grid, and uses it to prioritize continuation blocks consistent with the established chain before broader trial decoding. Together, the LZW decode, sub-block phase, and continuity checks based on MAD and consumption rate give GIF reassembly stronger structural and heuristic signals than entropy-coded JPEG, which is reflected in its substantially higher out-of-order recovery in Section~\ref{sec:performance}.

\subsection{ELF File Carving}

\begin{figure}[h]
    \centering
    \includegraphics[scale=0.5]{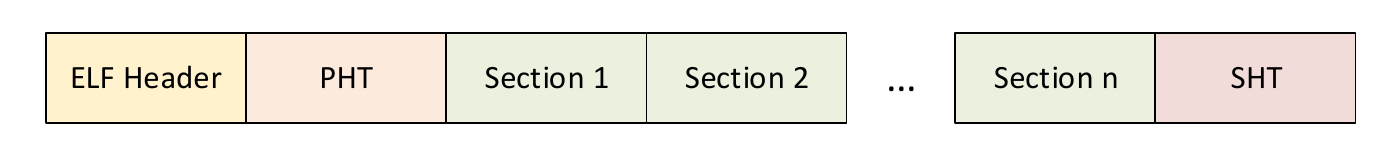}
    \caption{Simplified representation of the ELF file format.}
    \label{fig:elf-format}
\end{figure}

Executable and Linkable Format (ELF)~\cite{ELFSpec} files are the standard executable and object format on many UNIX-like systems, including Linux servers, embedded and IoT devices, and Android's native components. Recovering deleted executables matters in intrusion cases, where malware and attacker tooling are routinely deleted after use; ELF also illustrates a validator pattern that generalizes: length recovery from internal metadata for a format with no footer. An ELF begins with a fixed header followed by two key metadata tables: the \emph{program header table} (PHT), which describes loadable segments and their file offsets, and the \emph{section header table} (SHT), which describes on-disk sections. The bulk file contents (code, data, relocations, symbols, etc.) are organized into sections and/or segments, as illustrated in Figure~\ref{fig:elf-format}.

Unlike many container formats, ELF has no explicit footer. As a result, simple signature-based carvers can misestimate length and emit unrelated trailing bytes. Scalpel3 mitigates this by estimating the file span from internal ELF metadata. After locating and checking the ELF header, Scalpel3 derives the extents of the ELF header, program header table (PHT), and, when present, section header table (SHT). When an SHT is available, Scalpel3 computes the file end as the highest offset reached by these structures or by any section that occupies bytes in the file, excluding zero-size and \texttt{NOBITS} sections such as \texttt{.bss}. The estimate is rounded to a 4-byte boundary for 32-bit ELF files and an 8-byte boundary for 64-bit files. Because the calculation uses absolute file offsets, it accounts for padding between sections. These bounds generally identify the correct file length, allowing Scalpel3 to recover the complete ELF file. Trailing padding not described by an internal ELF structure can leave the estimated file length short by a few bytes.

The SHT is optional in ELF, so Scalpel3 cannot rely on it unconditionally. When the SHT is absent, Scalpel3 instead uses the highest offset reached by the ELF header, PHT, or any non-null segment that occupies bytes in the file. This provides a best-effort bound on the file span when section metadata are unavailable.

Once a size estimate is available, contiguous recovery is straightforward. After detecting the ELF magic bytes, Scalpel3 validates the full ELF header and then validates the metadata structures at their expected offsets (first the PHT and, if present, the SHT). If these checks pass, Scalpel3 carves the byte range from the header to the computed end offset.

Fragmented ELF recovery requires block-level evidence that metadata alone cannot supply: the bulk of an ELF file (machine code, data, and tables) has little per-block structure that a rule-based validator can check cheaply. The ELF block validator therefore uses a learned model~\cite{Waguespack2025}, invoked through the same ONNX Runtime integration as MoDiCo (Section~\ref{sec:machine-learning}), to classify the contents of candidate blocks during block validation.

\subsection{MP3 File Carving}

\begin{figure}[h]
    \centering
    \includegraphics[scale=0.35]{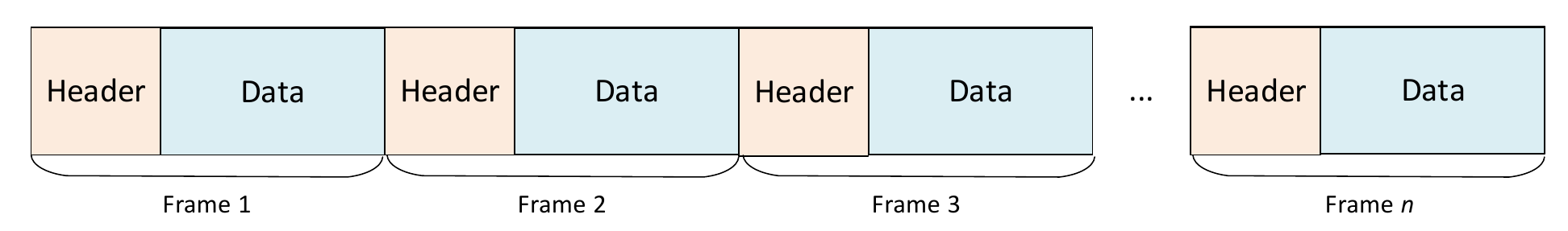}
    \caption{Simplified representation of the MP3 file format.}
    \label{fig:mp3-format}
\end{figure}

The MP3 format~\cite{MP3Spec} lacks the kind of explicit file-level header and footer used by many signature-based carvers. Instead, MP3 content is organized as a sequence of \emph{frames}, each consisting of a frame header followed by frame data, as depicted in Figure~\ref{fig:mp3-format}. Because traditional header--footer carving assumes an explicit start marker, MP3 carving requires a different notion of a ``header.'' When an optional container header such as ID3 is absent, Scalpel3 treats a block as a candidate MP3 start if it begins with a valid frame header and supports a sufficiently long chain of subsequent headers without error.

The frame header encodes enough information (e.g., bitrate, sampling rate, padding) to determine the expected length of the current frame and thus the predicted location of the next frame header. This structure enables detection via \emph{header chaining}: if the first four bytes satisfy MP3 frame header constraints and the derived frame length repeatedly lands on additional valid frame headers, confidence that the region is MP3 increases substantially.

In practice, however, header chaining alone is not sufficient to eliminate false positives: unrelated data can occasionally satisfy frame header constraints by chance. Scalpel3 therefore combines block-level detection with stricter file-level validation. Candidates with a consistent header chain receive high confidence. File validation then rejects impossible frame lengths, invalid parameter combinations, and inconsistent continuation points. If the final observed frame appears truncated by a filesystem block boundary, the validator can return \textit{promising} rather than \textit{validates}, indicating that the stream is internally consistent so far but incomplete within the current block.

Fragmented MP3 reassembly is more challenging because frame boundaries frequently cross filesystem block boundaries. Scalpel3 records boundary offsets when a frame extends past the end of a block (a \emph{rear} cutoff) and when a valid frame chain begins after the start of a block (a \emph{front} offset). These offsets provide lightweight join constraints: blocks whose rear and front offsets are compatible are plausible neighbors. When offsets are not unique, Scalpel3 uses optional frame CRC fields when applicable and otherwise compares spectral peaks derived from decoded audio to disambiguate joins. These techniques are developed further in~\cite{Hendrick2025MP3}.

\subsection{ZIP and Office Open XML File Carving}
\label{sec:zip-office}

The ZIP container~\cite{ZIPSpec} underlies both ordinary archives and modern Microsoft Office documents. An archive stores a sequence of entries, each a local file header followed by optionally compressed data, and ends with a central directory listing every entry and an end of central directory (EOCD) record that locates it. The Scalpel3 ZIP validator exploits this redundancy: after matching the local file header signature at a candidate start, it scans forward for the EOCD, walks the central directory, and cross-checks each central directory entry against the corresponding local file header. The EOCD also yields the exact file length, including any trailing archive comment. Because the central directory summarizes the entire archive, these checks give strong evidence that a candidate is complete and internally consistent.

Office Open XML documents (DOCX, PPTX, XLSX) are ZIP containers and are carved by the same validator. A post-processing step then searches each recovered archive for characteristic entry names: \texttt{word/document} for DOCX, \texttt{ppt/slides} for PPTX, and \texttt{xl/workbook} for XLSX. Files with these markers are renamed to the corresponding Office extension; everything else remains \texttt{.zip}. Classification happens after carving because it does not affect recovery itself, only the labeling of outputs.

\subsection{Using and Extending Scalpel3}
\label{sec:idea-to-experiment}

Scalpel3 is an operational forensic carving tool as well as a framework for recovery research. Its use and extension fall into three levels:

\begin{enumerate}
    \item \emph{Operational forensic use.} An investigator uses the existing file format support directly against an evidence image. This requires nothing beyond installing scalpel3.
    \item \emph{Adding standard format support.} A developer implements a file validator and registers the format. If the validator reports useful incremental progress as blocks are appended, the left-to-right (LR) baseline can reconstruct fragmented candidates. Block validation and state are added only when they provide useful evidence.
    \item \emph{Developing a custom recovery strategy.} When the LR baseline cannot efficiently exploit the format's structure, a researcher can add block evidence, checkpointable state, and a custom reassembly function. The PNG strategy described above illustrates this level.
\end{enumerate}

The PNG strategy of Section~\ref{sec:png-carving} illustrates the third level end to end. The recovery problem is fragmented PNGs whose fragments may be separated by gaps or stored out of order, with no filesystem metadata available. The format knowledge is the chunk grammar: length-prefixed, typed chunks, each protected by a CRC. That knowledge determines the block validator, which records each block's CRC32 as block state (Section~\ref{block-state}) during block validation. The file validator needs nothing beyond the standard interface: it reports \texttt{validates\_to} incrementally, so failing candidates are truncated and enqueued as promising by the ordinary machinery (Section~\ref{sec:adding-new-file-types}). The LR baseline can extend such candidates only by trial validation; it cannot exploit the stored CRCs, which is why the format warrants a custom reassembly function that solves for continuation blocks instead of enumerating them. Registration names these components, and no backend changes are involved.

When developing or evaluating new format support, the toolchain of Section~\ref{sec:scalpel3-toolchain} supports reproducible experiments: \texttt{fragmentator} generates a test image with specified fragmentation and a ground truth key file, \texttt{crblockmap} builds the image's blockmap, and \texttt{scalpel3} carves the image using that blockmap, with \texttt{scalpel3-ctl} available for monitoring. Outputs are organized by type and status and optionally include blockvector listings of the actual blocks used; recovered files are compared with the key file to determine whether they match the originals exactly, following the methodology of Section~\ref{sec:performance}.

\section{Performance}
\label{sec:performance}

This section evaluates Scalpel3's performance: contiguous recovery measured against PhotoRec, and fragmented recovery under three controlled scenarios: gaps (GAP), out-of-order block placement (OOO), and both together (GAP+OOO).

\subsection{Experimental Setup and Methodology}
\label{sec:perf-setup}

All performance experiments reported in this section were conducted under Ubuntu on an otherwise idle workstation equipped with a 96-core AMD Ryzen Threadripper Pro 7995WX CPU and 512~GiB of RAM.  The 96-core processor provides 192 hardware threads via simultaneous multithreading. We tested thread counts through 192 but report results through 128 because additional speedup beyond 128 was not consistent across workloads; we believe memory and cache bandwidth are exhausted before thread counts reach the number of logical cores. 

All experiments use the same corpus of 80{,}549 files: 52{,}630 PNG, 5{,}000 ZIP, 9{,}988 JPEG, 4{,}929 GIF, 4{,}997 ELF, and 3{,}005 DOCX. The 5{,}000 ZIP files are all multi-entry archives. Test images are generated from this corpus with \texttt{fragmentator} (Section~\ref{sec:scalpel3-toolchain}), which also emits the ground truth used for scoring. All test images use a 16~KiB carving block size. Recovered outputs are compared with ground truth; byte-for-byte matches are reported as complete, and the tables also distinguish partial, incorrect, and unrecovered files. Partial combines a correct block sequence with a length mismatch, a correct prefix, or a recovery whose only mismatched block is the terminal block. Incorrect means that the best output has a nonterminal block mismatch or is a full-length reconstruction that fails byte-for-byte verification.

Each configuration was executed twice: contiguous results report minimum and maximum observed runtimes, fragmented progress curves are averaged across the two runs, and recovery tables report both run values. Progress curves in fragmented reassembly tests count only files accepted by Scalpel3's validators, while the accompanying tables classify each ground truth file by its best recovered output.

PhotoRec 7.3-WIP, which is single-threaded, serves as the performance baseline for contiguous carving. It is a widely used open source carver, and to our knowledge no publicly available tool performs general fragmented reassembly (Section~\ref{sec:related-work}), so the comparison covers the contiguous workload both tools share. 

\begin{figure}[!htbp]
    \centering
    \includegraphics[width=\linewidth]{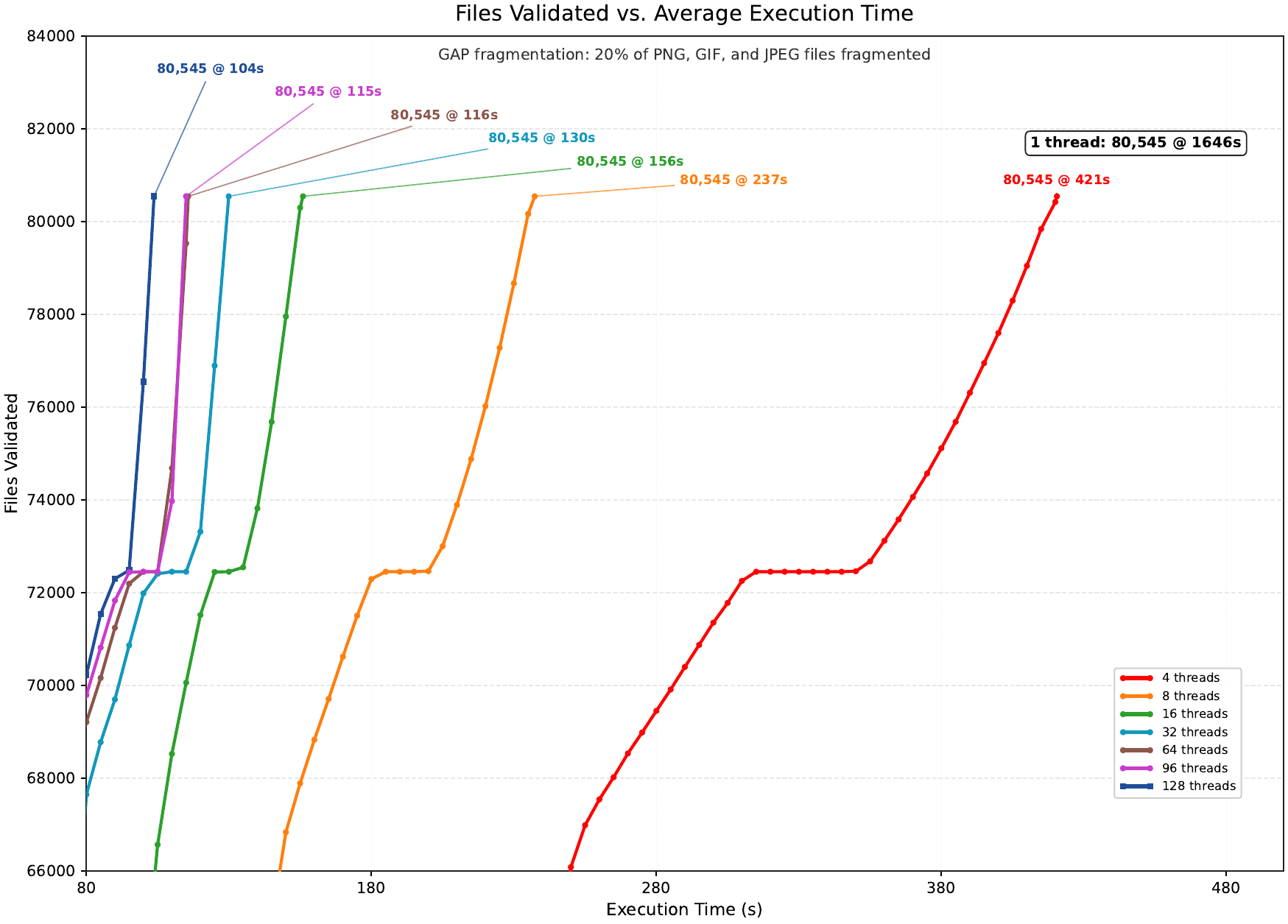}
    \caption{Average recovery progress for the GAP fragmentation scenario. Each curve averages two runs and is linearly interpolated between periodic status counters reporting files validated. Axes are cropped to 80--500 seconds and 66{,}000--84{,}000 files to make the multithreaded results legible; the single thread endpoint is reported separately because it lies beyond the displayed time range. Every configuration reaches the same final count of validated files and natural completion, while additional threads reduce the time required to do so.}
    \label{fig:gap20-progress}
\end{figure}

\begin{figure}[!htbp]
    \centering
    \includegraphics[width=\linewidth]{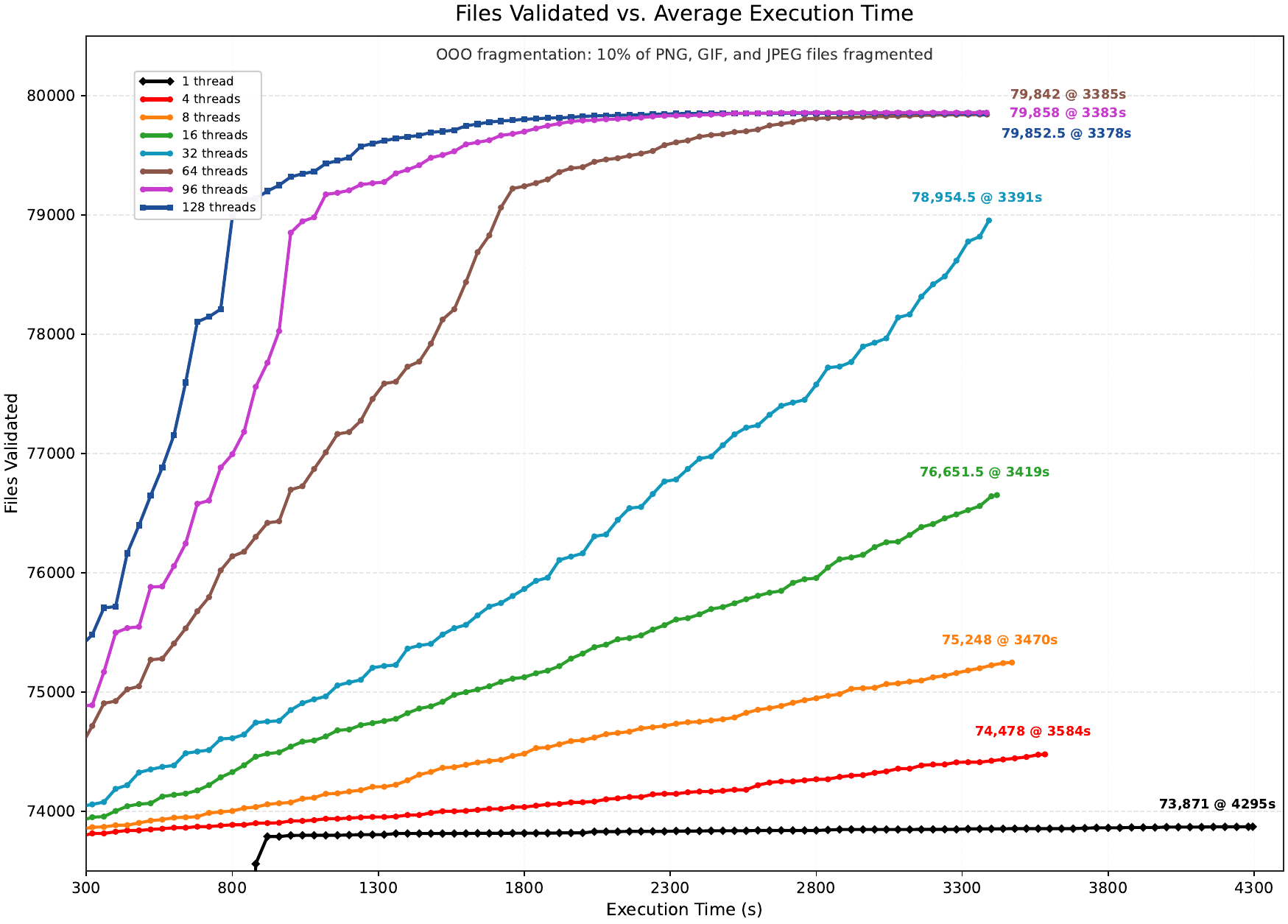}
    \caption{Average recovery progress for the out-of-order fragmentation scenario. Each curve averages two runs and is linearly interpolated between periodic status counters reporting files validated. Axes are cropped to 300--4{,}400 seconds and 73{,}500--80{,}500 files to make differences among thread counts legible. Higher thread counts reach the high recovery region much earlier, with the 64-, 96-, and 128-thread configurations converging near the end of the fixed reassembly budget.}
    \label{fig:ooo10-progress}
\end{figure}

\begin{figure}[!htbp]
    \centering
    \includegraphics[width=\linewidth]{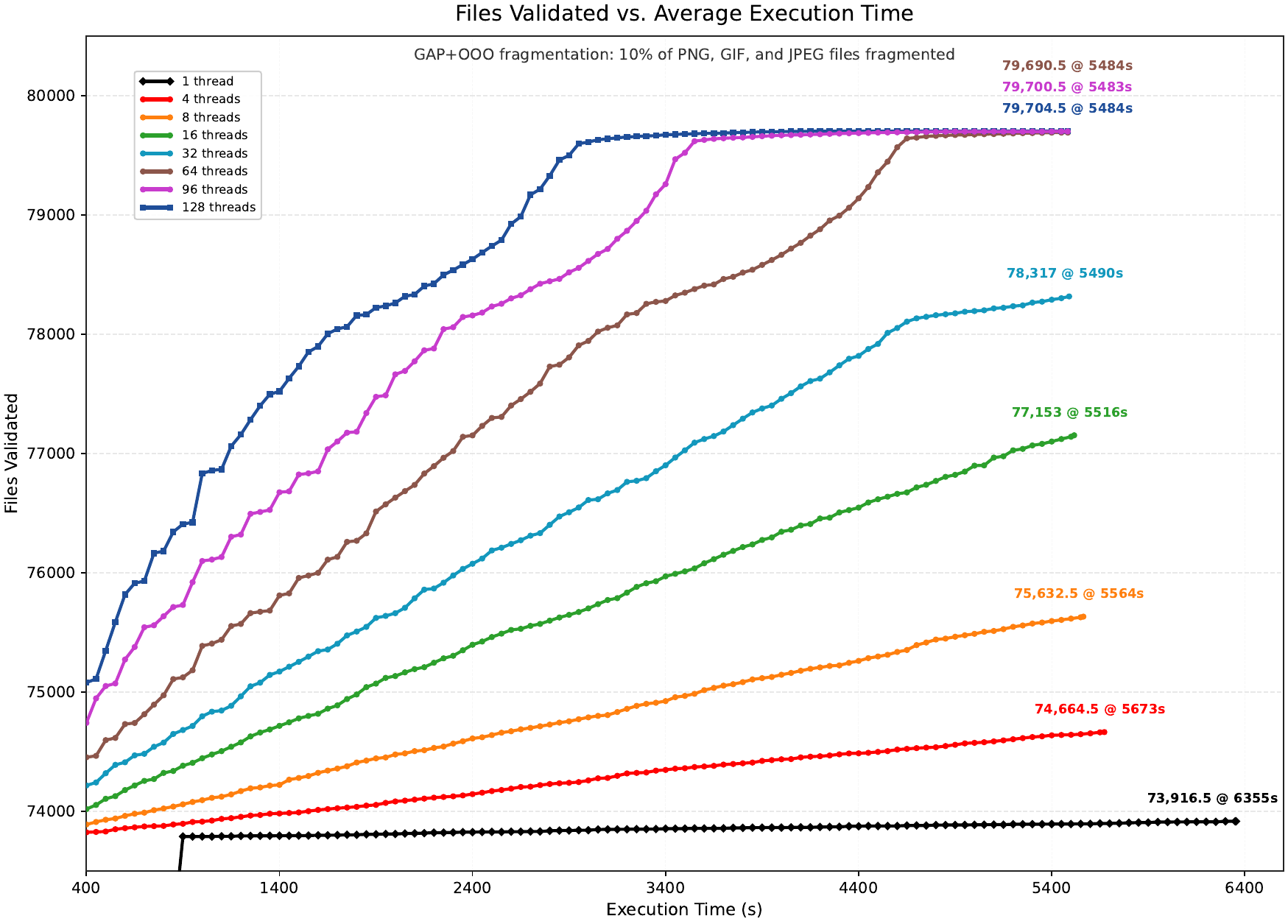}
    \caption{Average recovery progress for the combined GAP and out-of-order fragmentation scenario. Each curve averages two runs and is linearly interpolated between periodic status counters reporting files validated. Axes are cropped to 400--6{,}600 seconds and 73{,}500--80{,}500 files to make differences among thread counts legible. Higher thread counts reach the high recovery region earlier, while the 64-, 96-, and 128-thread configurations converge near the end of the fixed reassembly budget.}
    \label{fig:gap-ooo10-progress}
\end{figure}

The evaluation supports four principal findings:

\begin{itemize}
    \item Contiguous recovery reaches PhotoRec's performance range while recovering 80{,}524--80{,}525 files exactly, about 5{,}000 more than PhotoRec (Section~\ref{sec:perf-contiguous}, Figures~\ref{fig:contig-no-padding}--\ref{fig:contig-random10m}, and Table~\ref{tab:contiguous-accuracy}).
    \item For contiguous recovery, introduction of zero blocks at random locations has little effect, while introduction of random blocks adds modest scanning cost (Section~\ref{sec:perf-contiguous}, Figures~\ref{fig:contig-zero10m} and~\ref{fig:contig-random10m}).
    \item The massively threaded backend substantially accelerates fragmented recovery. GAP runtime falls from 1{,}646~s with one thread to 104~s with 128, while higher thread counts make validated results available much earlier in the harder OOO and GAP+OOO workloads (Section~\ref{sec:perf-fragmented}, Figures~\ref{fig:gap20-progress}--\ref{fig:gap-ooo10-progress}).
    \item Recovery accuracy reflects the structural evidence available in each format. PNG is recovered nearly perfectly, GIF remains strong, and JPEG is the limiting case under out-of-order fragmentation (Sections~\ref{sec:perf-gap}--\ref{sec:perf-gap-ooo}, Tables~\ref{tab:gap20-recovery}--\ref{tab:gap-ooo10-recovery}).
\end{itemize}

\subsection{Contiguous Recovery}
\label{sec:perf-contiguous}

This experiment evaluates contiguous carving performance against PhotoRec. At realistic multicore thread counts, Scalpel3 stays within PhotoRec's performance range and recovers more files exactly.

\begin{figure}[htbp]
    \centering
    \includegraphics[width=\linewidth]{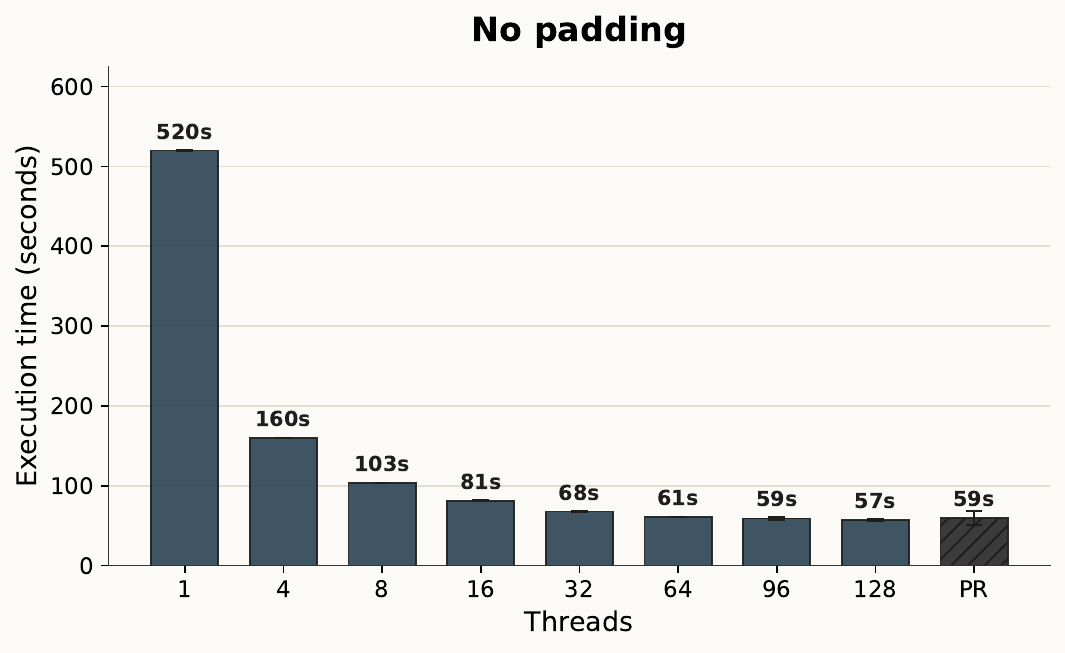}
    \caption{Contiguous recovery performance for the 80,549-file corpus with no additional padding. Scalpel3 is shown for thread counts 1--128; PR denotes single-threaded PhotoRec 7.3-WIP. Whiskers show minimum and maximum observed runtime.}
    \label{fig:contig-no-padding}
\end{figure}

\begin{figure}[htbp]
    \centering
    \includegraphics[width=\linewidth]{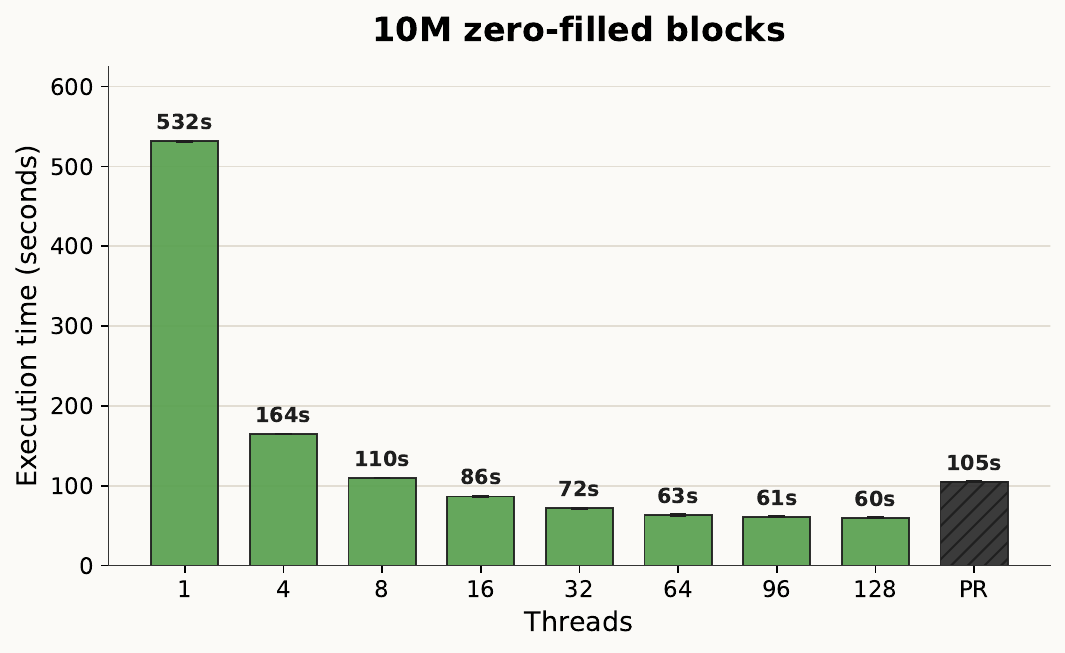}
    \caption{Contiguous recovery performance for the 80,549-file corpus with ten million zero-filled padding blocks. Scalpel3 is shown for thread counts 1--128; PR denotes single-threaded PhotoRec 7.3-WIP. Whiskers show minimum and maximum observed runtime.}
    \label{fig:contig-zero10m}
\end{figure}

\begin{figure}[htbp]
    \centering
    \includegraphics[width=\linewidth]{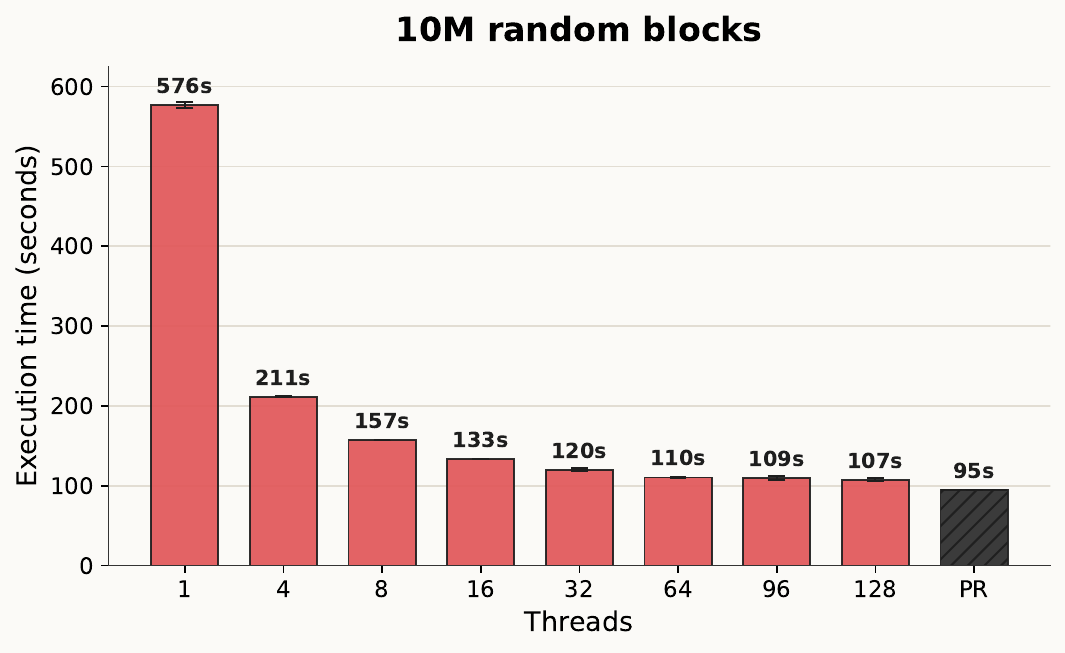}
    \caption{Contiguous recovery performance for the 80,549-file corpus with ten million random padding blocks. Scalpel3 is shown for thread counts 1--128; PR denotes single-threaded PhotoRec 7.3-WIP. Whiskers show minimum and maximum observed runtime.}
    \label{fig:contig-random10m}
\end{figure}

The test includes three scenarios. Each uses the same set of contiguous files, but differs in the additional blocks present in the disk image: (i) no additional blocks (a 103.33~GiB image), (ii) ten million zero-filled blocks inserted at random locations, and (iii) ten million random blocks inserted at random locations (255.92~GiB in both padded cases). Scalpel3 ran in contiguous-only mode with thread pool sizes of 1, 4, 8, 16, 32, 64, 96, and 128. Since this experiment is designed to evaluate contiguous file recovery, Scalpel3 was configured to write only fully validated files, and PhotoRec outputs were scored against the same ground truth.   Figures~\ref{fig:contig-no-padding}--\ref{fig:contig-random10m} show the results.

Scalpel3 uses the blockmap to avoid searching known all-zero blocks for headers and footers, while the random padding must be searched because it may contain signature bytes. With random padding, PhotoRec is faster in this contiguous-only workload, but Scalpel3 remains in the same performance range while doing setup work that PhotoRec does not attempt.

Across the three scenarios, Scalpel3 recovered 80{,}524--80{,}525 files exactly out of 80{,}549. PhotoRec recovered 75{,}434--75{,}540 files exactly. The difference is consistent with length estimation and overcarving behavior in some formats, particularly ELF binaries.

\begin{table}[htbp]
    \centering
    \footnotesize
    \setlength{\tabcolsep}{3.0pt}
    \begin{tabular}{@{}lrrrrrrr@{}}
        \toprule
        & & \multicolumn{2}{c}{No padding} & \multicolumn{2}{c}{Zero padding} & \multicolumn{2}{c}{Random padding} \\
        \cmidrule(lr){3-4}\cmidrule(lr){5-6}\cmidrule(l){7-8}
        Type & Files & Scalpel3 & PhotoRec & Scalpel3 & PhotoRec & Scalpel3 & PhotoRec \\
        \midrule
        DOCX & 3{,}005 & 3{,}005/0 & 3{,}003/2 & 3{,}005/0 & 2{,}992/13 & 3{,}005/0 & 3{,}003/2 \\
        ELF & 4{,}997 & 4{,}981/16 & 2/4{,}995 & 4{,}981/16 & 0/4{,}997 & 4{,}981/16 & 0/4{,}997 \\
        GIF & 4{,}929 & 4{,}928/1 & 4{,}924/5 & 4{,}928/1 & 4{,}927/2 & 4{,}928/1 & 4{,}927/2 \\
        JPEG & 9{,}988 & 9{,}988/0 & 9{,}985/3 & 9{,}987/1 & 9{,}986/2 & 9{,}988/0 & 9{,}986/2 \\
        PNG & 52{,}630 & 52{,}630/0 & 52{,}625/5 & 52{,}630/0 & 52{,}531/99 & 52{,}630/0 & 52{,}625/5 \\
        ZIP & 5{,}000 & 4{,}993/7 & 4{,}998/2 & 4{,}993/7 & 4{,}998/2 & 4{,}993/7 & 4{,}999/1 \\
        \midrule
        Total & 80{,}549 & 80{,}525/24 & 75{,}537/5{,}012 & 80{,}524/25 & 75{,}434/5{,}115 & 80{,}525/24 & 75{,}540/5{,}009 \\
        \bottomrule
    \end{tabular}
    \caption{Contiguous recovery accuracy by file type. Each entry reports exact/not exact ground truth files; exact recoveries match ground truth byte for byte, while not exact includes both outputs that failed external validation and files that were not recovered. Both repetitions produced identical counts.}
    \label{tab:contiguous-accuracy}
\end{table}

Table~\ref{tab:contiguous-accuracy} shows that Scalpel3 recovered every DOCX and PNG exactly in all three scenarios, as well as every JPEG except one in the zero-padding scenario. Its remaining shortfall consists of one GIF length mismatch, 16 partial ELF recoveries caused by length estimation, and seven ZIP files that were not recovered. PhotoRec's shortfall is dominated by ELF: at most two of the 4{,}997 ELF files matched ground truth exactly across any scenario.  Most of its other file types were recovered exactly, although the zero-padding scenario also left 99 PNG and 13 DOCX files without exact matches. 

\FloatBarrier
\subsection{Fragmented Recovery}
\label{sec:perf-fragmented}

We now evaluate fragmented recovery using three fragmentation models on the same 80{,}549-file
corpus: gap-based fragmentation (GAP), out-of-order fragmentation (OOO), and their combination
(GAP+OOO). We restrict fragmented evaluation to three representative formats: PNG, GIF, and JPEG. In every scenario, only these files are fragmented, while DOCX,
ELF, and ZIP files are kept contiguous. Each scenario embeds thousands of fragmented files among
millions of blocks of unrelated content. In GAP, unrelated data interrupts the logical block sequence
and divides each selected file into separate physical extents. In OOO, a contiguous run of file blocks
is removed from its logical position and stored elsewhere in the randomized image, requiring Scalpel3
to locate the displaced run and restore the correct ordering. GAP+OOO combines these
disruptions at different positions in the same file. Because out-of-order reassembly is substantially more expensive than bridging a gap (Scalpel3
must infer the correct block ordering rather than simply close a missing range), we fragment
20\% of the targeted files in the GAP scenario but 10\% in the OOO and GAP+OOO scenarios, so that
every thread configuration makes a reasonable amount of progress within a practical time budget for the experiments on a single workstation.

Each scenario was assigned a fragmented reassembly budget, measured from entry into that phase: 400~s for GAP, 3{,}300~s for OOO, and 5{,}400~s for GAP+OOO. The GAP runs all reached natural completion before exhausting their reassembly budget. For OOO and GAP+OOO, Scalpel3 stopped at the next safe checkpoint boundary after the budget expired and wrote a restart checkpoint. We set the budgets empirically by running the 128-thread configuration until its validated file count reached a substantial plateau, then choosing a practical stopping point that captures the bulk of achievable recovery. The progress figures use total elapsed time, including initialization, block validation, header/footer discovery, contiguous file validation, and fragmented reassembly. Their endpoints therefore include setup time in addition to the reassembly budget.

The three scenarios expose two complementary forms of scaling. GAP reaches the same final validated count at every thread count, so additional threads directly reduce time to completion. OOO and GAP+OOO retain difficult candidates at the end of the fixed budget; there, additional threads primarily improve time to result and the amount recovered within the budget. The discussion below considers both timing and accuracy, separated by file type and by whether each file was fragmented.

\FloatBarrier
\suppressfloats[t]
\subsubsection{Gap Fragmentation}
\label{sec:perf-gap}

The GAP scenario uses the same 80{,}549-file corpus. Specifically, 10{,}526 of 52{,}630 PNG files, 986 of 4{,}929 GIF files, and 1{,}998 of 9{,}988 JPEG files (20\% of each type) were fragmented using a gap-based layout, producing a 103.88~GiB image. This design isolates the behavior of Scalpel3's image reassembly strategies while preserving a large mixed-file workload.

Every thread configuration reaches the same final count of validated files and natural completion. Average total runtime decreases monotonically from 1{,}646~s with one thread to 421~s with four, 237~s with eight, 156~s with 16, 130~s with 32, 116~s with 64, 115~s with 96, and 104~s with 128. Computed from the unrounded run averages, the 128-thread result corresponds to a 15.9$\times$ speedup relative to one thread. Ground truth scoring finds 80{,}525 exact recoveries (Table~\ref{tab:gap20-recovery}).

\begin{table}[!htbp]
    \centering
    \footnotesize
    \setlength{\tabcolsep}{3.0pt}
    \begin{tabular}{@{}llrrrrr@{}}
        \toprule
        Type & Layout & Files & Complete & Partial & Incorrect & \shortstack{Not\\recovered} \\
        \midrule
        DOCX & Unfragmented & 3{,}005 & 3{,}005 & 0 & 0 & 0 \\
        ELF & Unfragmented & 4{,}997 & 4{,}981 & 16 & 0 & 0 \\
        GIF & Unfragmented & 3{,}943 & 3{,}942 & 1 & 0 & 0 \\
        GIF & Fragmented & 986 & 986 & 0 & 0 & 0 \\
        JPEG & Unfragmented & 7{,}990 & 7{,}990 & 0 & 0 & 0 \\
        JPEG & Fragmented & 1{,}998 & 1{,}998 & 0 & 0 & 0 \\
        PNG & Unfragmented & 42{,}104 & 42{,}104 & 0 & 0 & 0 \\
        PNG & Fragmented & 10{,}526 & 10{,}526 & 0 & 0 & 0 \\
        ZIP & Unfragmented & 5{,}000 & 4{,}993 & 0 & 0 & 7 \\
        \midrule
        \multicolumn{2}{l}{Corpus total} & 80{,}549 & 80{,}525 & 17 & 0 & 7 \\
        \bottomrule
    \end{tabular}
    \caption{Final recovery breakdown for the two GAP runs using 128 threads, separated by file type and layout. Both runs produced identical values.}
    \label{tab:gap20-recovery}
\end{table}
\FloatBarrier

Table~\ref{tab:gap20-recovery} provides the corresponding file type and layout breakdown for the 128-thread runs. All 13{,}510 fragmented files were recovered exactly: 10{,}526 PNG, 1{,}998 JPEG, and 986 GIF. Among the unfragmented files, every PNG and JPEG was exact, as were 3{,}942 of 3{,}943 GIF files; the remaining GIF had a length mismatch. The imperfect ELF and ZIP results are unrelated to fragmentation because those formats remain contiguous in this experiment. The ELF partials arise from length estimation: when an ELF contains trailing padding not described by any internal table, Scalpel3's size estimate can fall short by a few bytes (Section~\ref{sec:file-type-specific-carving}), yielding a length mismatch rather than an exact match. The seven unrecovered ZIP files reflect a current limitation of the ZIP validator on certain inputs.

\FloatBarrier
\suppressfloats[t]
\subsubsection{Out-of-Order Fragmentation}
\label{sec:perf-ooo}

This scenario uses the same corpus, but fragments 10\% of the PNG, GIF, and JPEG files by placing some file blocks out of logical order. Specifically, 5{,}263 PNG files, 493 GIF files, and 999 JPEG files were fragmented; the image is 103.33~GiB.

Out-of-order recovery makes the time-to-result benefit of parallelism clear. On average, the 128-thread configuration reaches 79{,}000 validated files after 806~s, compared with 1{,}045~s for 96 threads and 1{,}709~s for 64. By the end of the fixed reassembly budget, the 64-, 96-, and 128-thread configurations converge near 79{,}850 validated files, while lower thread counts remain behind. Accuracy separates the formats: PNG and GIF remain strong, while JPEG accounts for most incomplete files (Table~\ref{tab:ooo10-recovery}).

\begin{table}[!htbp]
    \centering
    \footnotesize
    \setlength{\tabcolsep}{3.0pt}
    \begin{tabular}{@{}llrrrrr@{}}
        \toprule
        Type & Layout & Files & Complete & Partial & Incorrect & \shortstack{Not\\recovered} \\
        \midrule
        DOCX & Unfragmented & 3{,}005 & 3{,}005 & 0 & 0 & 0 \\
        ELF & Unfragmented & 4{,}997 & 4{,}981 & 16 & 0 & 0 \\
        GIF & Unfragmented & 4{,}436 & 4{,}435 & 1 & 0 & 0 \\
        GIF & Fragmented & 493 & 490/489 & 2/3 & 1 & 0 \\
        JPEG & Unfragmented & 8{,}989 & 8{,}989 & 0 & 0 & 0 \\
        JPEG & Fragmented & 999 & 299/277 & 634/648 & 66/74 & 0 \\
        PNG & Unfragmented & 47{,}367 & 47{,}367 & 0 & 0 & 0 \\
        PNG & Fragmented & 5{,}263 & 5{,}252/5{,}253 & 10/9 & 1 & 0 \\
        ZIP & Unfragmented & 5{,}000 & 4{,}993 & 0 & 0 & 7 \\
        \midrule
        \multicolumn{2}{l}{Corpus total} & 80{,}549 & 79{,}811/79{,}789 & 663/677 & 68/76 & 7 \\
        \bottomrule
    \end{tabular}
    \caption{Final recovery breakdown for the two out-of-order runs using 128 threads, separated by file type and layout. Where runs differ, values are reported as run~1/run~2; single values were identical.}
    \label{tab:ooo10-recovery}
\end{table}
\FloatBarrier

Table~\ref{tab:ooo10-recovery} gives the corresponding accuracy breakdown. All unfragmented PNG and JPEG files were recovered exactly, as were 4{,}435 of 4{,}436 unfragmented GIF files. Among the fragmented files, 5{,}252--5{,}253 of 5{,}263 PNG files were exact, 9--10 were partial, one was incorrect, and 8--9 of the partials crossed the first fragmentation boundary. Of the 493 fragmented GIF files, 489--490 were exact, 2--3 were partial, and one was incorrect; none of the partials crossed the boundary. JPEG recovery was lower: 277--299 of the 999 fragmented files were exact, 634--648 were partial, and 66--74 were incorrect. Only three JPEG partials in each run crossed the first fragmentation boundary; most retained only the initial contiguous prefix. Every fragmented file produced at least a retained candidate. This difference is consistent with the formats: PNG and GIF provide stronger structural signals for reassembly, whereas JPEG entropy-coded data provides less deterministic evidence about arbitrary block order.

\FloatBarrier
\suppressfloats[t]
\subsubsection{Combined Gap and Out-of-Order Fragmentation}
\label{sec:perf-gap-ooo}

This scenario applies gap and out-of-order fragmentation to 10\% of the PNG, GIF, and JPEG files: 5{,}263 PNG files, 493 GIF files, and 999 JPEG files were fragmented; the image is 103.39~GiB. It is the hardest of the three scenarios, since the same file requires both bridging a gap and reordering displaced blocks.

The combined scenario produces the strongest separation among thread counts. On average, 128 threads reaches 78{,}000 validated files after 1{,}614~s, compared with 2{,}219~s for 96 threads, 3{,}054~s for 64, and 4{,}545~s for 32. The 64-, 96-, and 128-thread runs converge near 79{,}700 validated files by the end of the fixed reassembly budget, but the higher thread counts make most of those results available substantially earlier. The accuracy ordering repeats: PNG remains highly reliable, GIF remains strong, and JPEG is the limiting format (Table~\ref{tab:gap-ooo10-recovery}).

\FloatBarrier

\begin{table}[!htbp]
    \centering
    \footnotesize
    \setlength{\tabcolsep}{3.0pt}
    \begin{tabular}{@{}llrrrrr@{}}
        \toprule
        Type & Layout & Files & Complete & Partial & Incorrect & \shortstack{Not\\recovered} \\
        \midrule
        DOCX & Unfragmented & 3{,}005 & 3{,}005 & 0 & 0 & 0 \\
        ELF & Unfragmented & 4{,}997 & 4{,}981 & 16 & 0 & 0 \\
        GIF & Unfragmented & 4{,}436 & 4{,}435 & 1 & 0 & 0 \\
        GIF & Fragmented & 493 & 476/477 & 15/14 & 2 & 0 \\
        JPEG & Unfragmented & 8{,}989 & 8{,}989 & 0 & 0 & 0 \\
        JPEG & Fragmented & 999 & 192/186 & 564/574 & 243/239 & 0 \\
        PNG & Unfragmented & 47{,}367 & 47{,}367 & 0 & 0 & 0 \\
        PNG & Fragmented & 5{,}263 & 5{,}226/5{,}217 & 37/45 & 0/1 & 0 \\
        ZIP & Unfragmented & 5{,}000 & 4{,}993 & 0 & 0 & 7 \\
        \midrule
        \multicolumn{2}{l}{Corpus total} & 80{,}549 & 79{,}664/79{,}650 & 633/650 & 245/242 & 7 \\
        \bottomrule
    \end{tabular}
    \caption{Final recovery breakdown for the two combined GAP and out-of-order runs using 128 threads, separated by file type and layout. Where runs differ, values are reported as run~1/run~2; single values were identical.}
    \label{tab:gap-ooo10-recovery}
\end{table}
\FloatBarrier

Table~\ref{tab:gap-ooo10-recovery} gives the corresponding accuracy breakdown. All unfragmented PNG and JPEG files were recovered exactly, as were 4{,}435 of 4{,}436 unfragmented GIF files. Among the fragmented files, 5{,}217--5{,}226 of 5{,}263 PNG files were exact, 37--45 were partial, zero or one was incorrect, and 11--17 of the partials crossed the first fragmentation boundary. Of the 493 fragmented GIF files, 476--477 were exact, 14--15 were partial, and two were incorrect; one partial in each run crossed the boundary. JPEG again accounts for most of the shortfall: 186--192 of the 999 fragmented files were exact, 564--574 were partial, and 239--243 were incorrect. Only three JPEG partials in each run crossed the first fragmentation boundary. 

\FloatBarrier

\section{Limitations}
\label{sec:limitations}

Fragmented reassembly requires internal structure that a validator can check: checksums, chunk
grammars, decodable streams, or metadata with offsets and lengths. Formats without such structure,
and data that is encrypted or externally compressed, provide few constraints for reassembly. The
JPEG results in Section~\ref{sec:performance} show the effect: entropy-coded data supports far less
accurate reordering than CRC-protected PNG.

Support for each file type is a separate research effort. An effective fragmented recovery strategy
requires deep study of the format, including which structures can be validated cheaply and which
constraints can prune the search. The backend supplies thread scheduling, I/O, block management,
checkpointing, and other infrastructure, allowing researchers to focus on support for a specific file
type. It does not replace analysis of that format, and results for one format do not necessarily
transfer to another, but we believe that the requirement of writing only select single-threaded functions to experiment with support for new file types substantially reduces the required effort.

Highly speculative fragmented recovery can be computationally expensive. For damaged inputs that
contain many missing blocks, for pathological fragmentation, or with weak validators, the search space may be effectively
unbounded and there is no termination guarantee (Section~\ref{sec:perf-setup}). Scalpel3 mitigates
this with stopping criteria, restart checkpoints, operator control, and phase ordering that postpones
the most speculative candidates until more constrained work has been attempted.

\section{Future Work}
\label{sec:future-work}

An important direction is broadening fragmented recovery to additional file formats. Each format presents a distinct research problem: identifying structures, offsets, checksums, and cross-field constraints that can prune reassembly without rejecting valid fragments. Scalpel3's public validator and custom reassembly interfaces provide a common platform for developing these strategies. Scalpel3 is open source, and we welcome community contributions of new format support.

Beyond file type coverage, Scalpel3 provides a default left-to-right reassembly strategy and supports format-specific custom reassembly functions. Custom strategies already exploit the structure of their formats; for the many formats that rely on the default, it is an open question whether additional generic strategies, selectable at registration, would serve some of them better. Candidates include bidirectional assembly, seam-based scoring, beam search, and constraint-based placement; each could supplement the left-to-right strategy and be evaluated against it under identical corpus, validation, and scheduling conditions.

Another direction is filesystem-aware carving. Scalpel3 deliberately operates on the raw block sequence because carving is most useful when filesystem metadata is absent or untrustworthy. Even incomplete knowledge of the originating filesystem could sharpen fragmented recovery. Allocators leave recognizable patterns: ext4 tries to keep a file's blocks within the same block group, small NTFS data attributes can be resident in an MFT record, and nonresident file data is allocated in filesystem clusters~\cite{LinuxExt4Overview,MicrosoftNTFSAttributes}. Scalpel3 can use such evidence without architectural changes: block validators can record it, and reassembly heuristics can prefer gap sizes and displacement patterns consistent with the allocator. The blockmap toolchain already supports a coarse form of this cooperation, since filesystem-aware tools can mark allocated regions as covered before carving begins (Section~\ref{sec:scalpel3-toolchain}).

\section{Conclusions}
\label{sec:conclusions}

This paper has presented Scalpel3, an extensible, massively parallel framework for recovering both contiguous and fragmented files; to our knowledge, no other publicly available system combines these capabilities. The architecture separates three concerns: the carver orchestrates the recovery pipeline and manages thread pools; the file mirror handles all disk I/O, maintains the blockmap, and provides a progressively simplified apparent image; and format-specific validators and reassembly functions supply the file type knowledge needed to guide reconstruction. This separation enables researchers to add support for new file types by implementing format-specific validation logic and, when needed, block validators, state hooks, or custom reassembly functions. Researchers write these components as single-threaded code. The backend handles parallelism, synchronization, and I/O automatically.

The phased recovery model (C, F1, F2) prioritizes easy recoveries first and progressively escalates to more expensive fragmented reassembly. Blockmap updates between successful recovery passes and at periodic checkpoints ensure that later passes benefit from the reduced search space created by prior recoveries. Practical features support real-world investigative use beyond controlled experiments: persistent checkpointing, human-in-the-loop control via \texttt{scalpel3-ctl}, block-level deduplication, and a FUSE filesystem for hybrid workflows.

Our performance results show high throughput for contiguous recovery and substantial acceleration of fragmented reassembly, with higher thread counts recovering more validated files earlier from large candidate sets. By releasing Scalpel3 as open source, we aim to provide a shared platform that lowers the barrier to experimentation and helps translate file carving research into practical tools.

\bibliographystyle{elsarticle-num}

\bibliography{Bib}

@inproceedings{b1,
  author       = {{Richard III}, Golden G. and Vassil Roussev},
  title        = {{{Scalpel}: A Frugal, High Performance File Carver}},
  booktitle    = {{Proceedings of the Fifth Annual Digital Forensics Research Workshop (DFRWS USA 2005)}},
  year         = {2005},
  month        = aug,
  address      = {New Orleans, LA}
}

@article{b3,
  author  = {Scott Hand and Zhiqiang Lin and Guofei Gu and Bhavani Thuraisingham},
  title   = {{{Bin-Carver}: Automatic Recovery of Binary Executable Files}},
  journal = {{Digital Investigation}},
  volume  = {9},
  pages   = {S108--S117},
  year    = {2012},
  doi     = {10.1016/j.diin.2012.05.014}
}

@misc{onnxruntime,
  title        = {{ONNX Runtime}},
  author       = {{ONNX Runtime Developers}},
  year         = {2026},
  howpublished = {\url{https://onnxruntime.ai/}},
  note         = {{Version} 1.26.0. Accessed: July 30, 2026}
}

@misc{LinuxExt4Overview,
  author       = {{Linux Kernel Documentation}},
  title        = {{ext4 Data Structures and Algorithms: High Level Design}},
  year         = {2026},
  howpublished = {\url{https://docs.kernel.org/filesystems/ext4/overview.html}},
  note         = {Accessed: August 5, 2026}
}

@misc{MicrosoftNTFSAttributes,
  author       = {{Microsoft}},
  title        = {{{ATTRIBUTE\_RECORD\_HEADER} Structure}},
  year         = {2026},
  howpublished = {\url{https://learn.microsoft.com/en-us/windows/win32/devnotes/attribute-record-header}},
  note         = {Accessed: August 5, 2026}
}

@article{b4,
  author  = {Simson L. Garfinkel},
  title   = {{Carving Contiguous and Fragmented Files with Fast Object Validation}},
  journal = {{Digital Investigation}},
  volume  = {4},
  pages   = {S2--S12},
  year    = {2007},
  doi     = {10.1016/j.diin.2007.06.017}
}

@misc{DFRWS2007ChallengeRepo,
  author       = {{Digital Forensics Research Workshop (DFRWS)}},
  title        = {{DFRWS 2007 Forensics Challenge}},
  howpublished = {\url{https://github.com/dfrws/dfrws2007-challenge}},
  year         = {2007},
  note         = {Accessed: July 30, 2026}
}

@misc{DFRWS2006ChallengeRepo,
  author       = {{Digital Forensics Research Workshop (DFRWS)}},
  title        = {{DFRWS 2006 Forensics Challenge}},
  howpublished = {\url{https://github.com/dfrws/dfrws2006-challenge}},
  year         = {2006},
  note         = {Accessed: July 30, 2026}
}

@article{MarzialeRichardRoussev2007GPU,
  author  = {Lodovico Marziale and {Richard III}, Golden G. and Vassil Roussev},
  title   = {{Massive Threading: Using {GPUs} to Increase the Performance of Digital Forensics Tools}},
  journal = {{Digital Investigation}},
  volume  = {4},
  pages   = {S73--S81},
  year    = {2007},
  doi     = {10.1016/j.diin.2007.06.014}
}

@incollection{RichardRoussevMarziale2007InPlace,
  author    = {{Richard III}, Golden G. and Vassil Roussev and Lodovico Marziale},
  title     = {{In-Place File Carving}},
  booktitle = {{Advances in Digital Forensics III}},
  editor    = {Philip Craiger and Sujeet Shenoi},
  series    = {{IFIP International Federation for Information Processing}},
  volume    = {242},
  publisher = {Springer},
  address   = {New York, NY},
  pages     = {217--230},
  year      = {2007},
  doi       = {10.1007/978-0-387-73742-3_15}
}

@misc{photorec,
  author       = {Christophe Grenier},
  title        = {{{PhotoRec}: Digital Picture and File Recovery}},
  howpublished = {\url{https://www.cgsecurity.org/wiki/PhotoRec}},
  note         = {{Version} 7.3-WIP. Free and open-source signature-based file carver. Accessed: July 30, 2026},
  year         = {2002},
  organization = {{CGSecurity}}
}

@article{Poisel2011Multimedia,
  author  = {Rainer Poisel and Simon Tjoa and Paul Tavolato},
  title   = {{Advanced File Carving Approaches for Multimedia Files}},
  journal = {{Journal of Wireless Mobile Networks, Ubiquitous Computing, and Dependable Applications}},
  volume  = {2},
  number  = {4},
  pages   = {42--58},
  year    = {2011},
  doi     = {10.22667/JOWUA.2011.12.31.042}
}

@article{Durmus2019JPEG,
  author  = {Emre Durmus and Pawe{\l} Korus and Nasir Memon},
  title   = {{Every Shred Helps: Assembling Evidence from Orphaned {JPEG} Fragments}},
  journal = {{IEEE Transactions on Information Forensics and Security}},
  volume  = {14},
  number  = {9},
  pages   = {2372--2386},
  year    = {2019},
  doi     = {10.1109/TIFS.2019.2897912}
}

@article{PalMemon2006,
  author  = {Nasir Memon and Anandabrata Pal},
  title   = {{Automated Reassembly of File Fragmented Images Using Greedy Algorithms}},
  journal = {{IEEE Transactions on Image Processing}},
  volume  = {15},
  number  = {2},
  pages   = {385--393},
  year    = {2006},
  doi     = {10.1109/TIP.2005.863054}
}

@article{PalSencarMemon2008SHT,
  author  = {Anandabrata Pal and Husrev T. Sencar and Nasir Memon},
  title   = {{Detecting File Fragmentation Point Using Sequential Hypothesis Testing}},
  journal = {{Digital Investigation}},
  year    = {2008},
  volume  = {5},
  pages   = {S2--S13},
  note    = {{Proceedings} of DFRWS 2008 USA},
  doi     = {10.1016/j.diin.2008.05.015}
}

@techreport{DHSAdroit2014,
  author      = {{U.S. Department of Homeland Security, Science and Technology Directorate}},
  title       = {{Test Results for Graphic File Carving Tool: {Adroit Photo Forensics 2013 v3.1d}}},
  institution = {{National Institute of Standards and Technology, Office of Law Enforcement Standards}},
  year        = {2014},
  month       = jul,
  note        = {{Report} dated July 16, 2014. \url{https://www.dhs.gov/sites/default/files/publications/508_Test%20Report_NIST_%20Adroit%20Photo%20Forensics%202013%20v3.1d_August%202015_Final_0.pdf}}
}

@misc{CarvFS2009,
  author       = {{Dutch National Police Agency (KLPD)}},
  title        = {{The Carve Path Zero-Storage Carving Library and Filesystem ({CarvFS} and {LibCarvPath})}},
  year         = {2006},
  howpublished = {\url{https://ocfa.sourceforge.net/libcarvpath/}},
  note         = {Accessed: July 30, 2026}
}

@inproceedings{Veenman2007ClusterClassification,
  author       = {Cor J. Veenman},
  title        = {{Statistical Disk Cluster Classification for File Carving}},
  booktitle    = {{Proceedings of the Third International Symposium on Information Assurance and Security (IAS 2007)}},
  pages        = {393--398},
  year         = {2007},
  organization = {IEEE},
  doi          = {10.1109/IAS.2007.75}
}

@article{Calhoun2008Fragments,
  author  = {William C. Calhoun and Drue Coles},
  title   = {{Predicting the Types of File Fragments}},
  journal = {{Digital Investigation}},
  volume  = {5},
  pages   = {S14--S20},
  year    = {2008},
  doi     = {10.1016/j.diin.2008.05.005}
}

@article{ByteRCNN2023,
  author  = {Kristian Skra{\v{c}}i{\'c} and Juraj Petrovi{\'c} and Predrag Pale},
  title   = {{{ByteRCNN}: Enhancing File Fragment Type Identification with Recurrent and Convolutional Neural Networks}},
  journal = {{IEEE Access}},
  volume  = {11},
  pages   = {138176--138187},
  year    = {2023},
  doi     = {10.1109/ACCESS.2023.3340441}
}

@inproceedings{mcdaniel2003content,
  title        = {{Content Based File Type Detection Algorithms}},
  author       = {Mason McDaniel and Mohammad Hossain Heydari},
  booktitle    = {{Proceedings of the 36th Annual Hawaii International Conference on System Sciences (HICSS 2003), Track 9}},
  year         = {2003},
  organization = {IEEE},
  doi          = {10.1109/HICSS.2003.1174905}
}

@incollection{karresand2006oscar,
  title     = {{{Oscar}---File Type Identification of Binary Data in Disk Clusters and {RAM} Pages}},
  author    = {Martin Karresand and Nahid Shahmehri},
  booktitle = {{Security and Privacy in Dynamic Environments}},
  editor    = {Simone Fischer-H{\"u}bner and Kai Rannenberg and Louise Yngstr{\"o}m and Stefan Lindskog},
  series    = {{IFIP International Federation for Information Processing}},
  volume    = {201},
  pages     = {413--424},
  year      = {2006},
  publisher = {Springer},
  address   = {Boston, MA},
  doi       = {10.1007/0-387-33406-8_35}
}

@article{axelsson2010normalised,
  title     = {{The Normalised Compression Distance as a File Fragment Classifier}},
  author    = {Stefan Axelsson},
  journal   = {{Digital Investigation}},
  volume    = {7},
  pages     = {S24--S31},
  year      = {2010},
  doi       = {10.1016/j.diin.2010.05.004}
}

@article{fitzgerald2012using,
  title     = {{Using {NLP} Techniques for File Fragment Classification}},
  author    = {Simran Fitzgerald and George Mathews and Colin Morris and Oles Zhulyn},
  journal   = {{Digital Investigation}},
  volume    = {9},
  pages     = {S44--S49},
  year      = {2012},
  doi       = {10.1016/j.diin.2012.05.008}
}

@incollection{sportiello2012context,
  title     = {{Context-Based File Block Classification}},
  author    = {Luigi Sportiello and Stefano Zanero},
  booktitle = {{Advances in Digital Forensics VIII}},
  editor    = {Gilbert Peterson and Sujeet Shenoi},
  series    = {{IFIP Advances in Information and Communication Technology}},
  volume    = {383},
  pages     = {67--82},
  year      = {2012},
  publisher = {Springer},
  address   = {Berlin, Heidelberg},
  doi       = {10.1007/978-3-642-33962-2_5}
}

@article{beebe2013sceadan,
  title     = {{{Sceadan}: Using Concatenated {N}-Gram Vectors for Improved File and Data Type Classification}},
  author    = {Nicole L. Beebe and Laurence A. Maddox and Lishu Liu and Minghe Sun},
  journal   = {{IEEE Transactions on Information Forensics and Security}},
  volume    = {8},
  number    = {9},
  pages     = {1519--1530},
  year      = {2013},
  doi       = {10.1109/TIFS.2013.2274728}
}

@inproceedings{wang2018ftcnn,
  author    = {Yanchao Wang and Zhongqian Su and Dayi Song},
  title     = {{File Fragment Type Identification with Convolutional Neural Networks}},
  booktitle = {{Proceedings of the 2018 International Conference on Machine Learning Technologies}},
  year      = {2018},
  pages     = {41--47},
  publisher = {Association for Computing Machinery},
  doi       = {10.1145/3231884.3231889}
}

@inproceedings{chen2018filefragment,
  author       = {Qian Chen and Qing Liao and Zoe L. Jiang and Junbin Fang and Siuming Yiu and Guikai Xi and Rong Li and Zhengzhong Yi and Xuan Wang and Lucas C. K. Hui and Dong Liu and En Zhang},
  title        = {{File Fragment Classification Using Grayscale Image Conversion and Deep Learning in Digital Forensics}},
  booktitle    = {{2018 IEEE Security and Privacy Workshops (SPW)}},
  year         = {2018},
  pages        = {140--147},
  organization = {IEEE},
  doi          = {10.1109/SPW.2018.00029}
}

@incollection{saaim2022light,
  title     = {{Light-Weight File Fragments Classification Using Depthwise Separable Convolutions}},
  author    = {Kunwar Muhammed Saaim and Muhamad Felemban and Saleh Alsaleh and Ahmad Almulhem},
  booktitle = {{ICT Systems Security and Privacy Protection}},
  editor    = {Weizhi Meng and Simone Fischer-H{\"u}bner and Christian D. Jensen},
  series    = {{IFIP Advances in Information and Communication Technology}},
  volume    = {648},
  pages     = {196--211},
  year      = {2022},
  publisher = {Springer},
  address   = {Cham},
  doi       = {10.1007/978-3-031-06975-8_12}
}

@article{haque2022byte2vec,
  title     = {{Byte Embeddings for File Fragment Classification}},
  author    = {Md Enamul Haque and Mehmet Engin Tozal},
  journal   = {{Future Generation Computer Systems}},
  volume    = {127},
  pages     = {448--461},
  year      = {2022},
  doi       = {10.1016/j.future.2021.09.019}
}

@article{mittal2020fifty,
  title     = {{{FiFTy}: Large-Scale File Fragment Type Identification Using Convolutional Neural Networks}},
  author    = {Govind Mittal and Pawe{\l} Korus and Nasir Memon},
  journal   = {{IEEE Transactions on Information Forensics and Security}},
  volume    = {16},
  pages     = {28--41},
  year      = {2021},
  doi       = {10.1109/TIFS.2020.3004266}
}

@article{bhatt2020hierarchy,
  title     = {{Hierarchy-Based File Fragment Classification}},
  author    = {Manish Bhatt and Avdesh Mishra and Md Wasi Ul Kabir and S. E. Blake-Gatto and Rishav Rajendra and Md Tamjidul Hoque and Irfan Ahmed},
  journal   = {{Machine Learning and Knowledge Extraction}},
  volume    = {2},
  number    = {3},
  pages     = {216--232},
  year      = {2020},
  doi       = {10.3390/make2030012}
}

@misc{FUSE,
  author       = {{The FUSE Project}},
  title        = {{FUSE: Filesystem in Userspace}},
  howpublished = {\url{https://github.com/libfuse/libfuse}},
  year         = {2026},
  note         = {Interface for userspace programs to export a filesystem to the Linux kernel. Accessed: July 30, 2026}
}

@misc{Kornblum2002Foremost,
  author       = {Kris Kendall and Jesse Kornblum},
  title        = {{Foremost}},
  howpublished = {\url{https://foremost.sourceforge.net/}},
  year         = {2001},
  note         = {Open source data carving tool originally developed at the U.S. Air Force Office of Special Investigations. Accessed: July 30, 2026}
}

@misc{PNGSpec,
  author       = {{World Wide Web Consortium}},
  title        = {{Portable Network Graphics ({PNG}) Specification (Second Edition)}},
  howpublished = {\url{https://www.w3.org/TR/2003/REC-PNG-20031110/}},
  year         = {2003},
  month        = nov,
  note         = {{W3C} Recommendation; {ISO/IEC} 15948:2003}
}

@misc{ELFSpec,
  author       = {{TIS Committee}},
  title        = {{Tool Interface Standard ({TIS}) Executable and Linking Format ({ELF}) Specification, Version 1.2}},
  year         = {1995},
  month        = may,
  howpublished = {\url{https://refspecs.linuxbase.org/elf/elf.pdf}}
}

@misc{MP3Spec,
  author       = {{International Organization for Standardization}},
  title        = {{{ISO/IEC 11172-3:1993}---Information Technology---Coding of Moving Pictures and Associated Audio for Digital Storage Media at up to about 1,5 {Mbit/s}---Part 3: Audio}},
  year         = {1993},
  month        = aug,
  howpublished = {\url{https://www.iso.org/standard/22412.html}},
  note         = {Edition 1}
}

@misc{GIFSpec,
  author       = {{CompuServe Incorporated}},
  title        = {{Graphics Interchange Format ({GIF}) Specification, Version 89a}},
  year         = {1990},
  month        = jul,
  howpublished = {\url{https://www.w3.org/Graphics/GIF/spec-gif89a.txt}}
}

@misc{ZIPSpec,
  author       = {{PKWARE, Inc.}},
  title        = {{{APPNOTE.TXT}---{.ZIP} File Format Specification, Version 6.3.10}},
  year         = {2022},
  month        = nov,
  howpublished = {\url{https://pkware.cachefly.net/webdocs/casestudies/APPNOTE.TXT}}
}

@article{HILGERT2019S22,
  author  = {Jan-Niclas Hilgert and Martin Lambertz and Mariia Rybalka and Roman Schell},
  title   = {{Syntactical Carving of {PNGs} and Automated Generation of Reproducible Datasets}},
  journal = {{Digital Investigation}},
  volume  = {29},
  pages   = {S22--S30},
  year    = {2019},
  doi     = {10.1016/j.diin.2019.04.014}
}

@article{VANDERMEER2024301687,
  author  = {Vincent {van der Meer} and Jeroen {van den Bos} and Hugo Jonker and Laurent Dassen},
  title   = {{Problem Solved: A Reliable, Deterministic Method for {JPEG} Fragmentation Point Detection}},
  journal = {{Forensic Science International: Digital Investigation}},
  volume  = {48},
  pages   = {301687},
  year    = {2024},
  note    = {{DFRWS} EU 2024 - Selected Papers from the 11th Annual Digital Forensics Research Conference Europe},
  doi     = {10.1016/j.fsidi.2023.301687}
}

@article{TANG2016S108,
  author  = {Yanbin Tang and Junbin Fang and K. P. Chow and S. M. Yiu and Jun Xu and Bo Feng and Qiong Li and Qi Han},
  title   = {{Recovery of Heavily Fragmented {JPEG} Files}},
  journal = {{Digital Investigation}},
  volume  = {18},
  pages   = {S108--S117},
  year    = {2016},
  doi     = {10.1016/j.diin.2016.04.016}
}

@article{SencarMemon2009,
  author  = {Husrev T. Sencar and Nasir Memon},
  title   = {{Identification and Recovery of JPEG Files with Missing Fragments}},
  journal = {{Digital Investigation}},
  volume  = {6},
  pages   = {S88--S98},
  year    = {2009},
  doi     = {10.1016/j.diin.2009.06.007}
}

@article{UzunSencar2015,
  author  = {Erkam Uzun and Husrev Taha Sencar},
  title   = {{Carving Orphaned JPEG File Fragments}},
  journal = {{IEEE Transactions on Information Forensics and Security}},
  volume  = {10},
  number  = {8},
  pages   = {1549--1563},
  year    = {2015},
  doi     = {10.1109/TIFS.2015.2416685}
}

@article{DeBockDeSmet2016,
  author  = {Johan {De Bock} and Patrick {De Smet}},
  title   = {{JPGcarve: An Advanced Tool for Automated Recovery of Fragmented JPEG Files}},
  journal = {{IEEE Transactions on Information Forensics and Security}},
  volume  = {11},
  number  = {1},
  pages   = {19--34},
  year    = {2016},
  doi     = {10.1109/TIFS.2015.2475238}
}

@article{UzunSencar2020,
  author  = {Erkam Uzun and Husrev Taha Sencar},
  title   = {{JpgScraper: An Advanced Carver for JPEG Files}},
  journal = {{IEEE Transactions on Information Forensics and Security}},
  volume  = {15},
  pages   = {1846--1857},
  year    = {2020},
  doi     = {10.1109/TIFS.2019.2953382}
}

@article{Cohen2007Advanced,
  author  = {M. I. Cohen},
  title   = {{Advanced Carving Techniques}},
  journal = {{Digital Investigation}},
  volume  = {4},
  number  = {3--4},
  pages   = {119--128},
  year    = {2007},
  doi     = {10.1016/j.diin.2007.10.001}
}

@article{GladyshevJames2017,
  author  = {Pavel Gladyshev and Joshua I. James},
  title   = {{Decision-Theoretic File Carving}},
  journal = {{Digital Investigation}},
  volume  = {22},
  pages   = {46--61},
  year    = {2017},
  doi     = {10.1016/j.diin.2017.08.001}
}

@article{Odogwu2020PNG,
  author  = {Kingson Chinedu Odogwu and Pavel Gladyshev and Babak Habibnia},
  title   = {{PNG Data Detector for DECA}},
  journal = {{Forensic Science International: Digital Investigation}},
  volume  = {32},
  pages   = {300910},
  year    = {2020},
  doi     = {10.1016/j.fsidi.2020.300910}
}

@article{Schneider2020LAYR,
  author  = {Janine Schneider and Hans-Peter Deifel and Stefan Milius and Felix Freiling},
  title   = {{Unifying Metadata-Based Storage Reconstruction and Carving with LAYR}},
  journal = {{Forensic Science International: Digital Investigation}},
  volume  = {33},
  pages   = {301006},
  year    = {2020},
  doi     = {10.1016/j.fsidi.2020.301006}
}

@article{Wang2024Fusion,
  author  = {Yi Wang and Wenyang Liu and Kejun Wu and Kim-Hui Yap and Lap-Pui Chau},
  title   = {{Intra- and Inter-Sector Contextual Information Fusion with Joint Self-Attention for File Fragment Classification}},
  journal = {{Knowledge-Based Systems}},
  volume  = {291},
  pages   = {111565},
  year    = {2024},
  doi     = {10.1016/j.knosys.2024.111565}
}

@article{Guzhov2025Transformer,
  author  = {Andrey Guzhov and Christoph Tobias Wirth},
  title   = {{Transformer-Based File Fragment Type Classification for File Carving in Digital Forensics}},
  journal = {{Proceedings of the European Conference on Cyber Warfare and Security}},
  volume  = {24},
  number  = {1},
  pages   = {169--176},
  year    = {2025},
  doi     = {10.34190/eccws.24.1.3552}
}

@misc{Scalpel3Software,
  author       = {{Richard III}, Golden G.},
  title        = {{Scalpel3}},
  year         = {2026},
  howpublished = {\url{https://github.com/nolaforensix/scalpel3-release}},
  note         = {Open source under the GPLv3; more than 100,000 lines of C, excluding third-party libraries; runs on macOS and Linux. Accessed: August 3, 2026}
}

@article{VanDerMeer2021Fragmentation,
  author  = {Vincent {van der Meer} and Hugo Jonker and Jeroen {van den Bos}},
  title   = {{A Contemporary Investigation of NTFS File Fragmentation}},
  journal = {{Forensic Science International: Digital Investigation}},
  volume  = {38},
  pages   = {301125},
  year    = {2021},
  doi     = {10.1016/j.fsidi.2021.301125}
}

@incollection{Huijsmans2026OOO,
  author    = {Nick Huijsmans and Bart Kuijsten and Hugo Jonker and Harm {van Beek}},
  title     = {{How to Carve Out-of-Order Fragmented Files}},
  booktitle = {{Juggling Formal Methods and Security: Essays Dedicated to Sjouke Mauw on the Occasion of His 65th Birthday}},
  editor    = {Barbara Fila and Hugo Jonker and Sa{\v{s}}a Radomirovi{\'c}},
  series    = {{Lecture Notes in Computer Science}},
  volume    = {16365},
  publisher = {Springer},
  pages     = {106--121},
  year      = {2026},
  doi       = {10.1007/978-3-032-20684-8_7}
}

@misc{ReviveIt2007,
  author       = {Joachim Metz and Robert-Jan Mora},
  title        = {{ReviveIt}},
  year         = {2007},
  howpublished = {\url{https://github.com/libyal/reviveit}},
  note         = {Proof of concept SmartCarving file recovery tool from the DFRWS 2006 and 2007 challenges; available as 2007 alpha source. Accessed: August 3, 2026}
}

@incollection{Lambertz2014Resurrection,
  author    = {Martin Lambertz and Rafael Uetz and Elmar Gerhards-Padilla},
  title     = {{Resurrection: A Carver for Fragmented Files}},
  booktitle = {{Digital Forensics and Cyber Crime (ICDF2C 2013)}},
  editor    = {Pavel Gladyshev and Andrew Marrington and Ibrahim Baggili},
  series    = {{Lecture Notes of the Institute for Computer Sciences, Social Informatics and Telecommunications Engineering}},
  volume    = {132},
  publisher = {Springer},
  pages     = {51--66},
  year      = {2014},
  doi       = {10.1007/978-3-319-14289-0_5}
}

@incollection{VanDenBos2012Excavator,
  author    = {Jeroen {van den Bos} and Tijs {van der Storm}},
  title     = {{Domain-Specific Optimization in Digital Forensics}},
  booktitle = {{Theory and Practice of Model Transformations (ICMT 2012)}},
  editor    = {Zhenjiang Hu and Juan {de Lara}},
  series    = {{Lecture Notes in Computer Science}},
  volume    = {7307},
  publisher = {Springer},
  pages     = {121--136},
  year      = {2012},
  doi       = {10.1007/978-3-642-30476-7_8}
}

@inproceedings{Qiu2014Carving,
  author    = {Weidong Qiu and Run Zhu and Jie Guo and Xiaoming Tang and Bozhong Liu and Zheng Huang},
  title     = {{A New Approach to Multimedia Files Carving}},
  booktitle = {{2014 IEEE International Conference on Bioinformatics and Bioengineering (BIBE)}},
  pages     = {105--110},
  year      = {2014},
  organization = {IEEE},
  doi       = {10.1109/BIBE.2014.31}
}

@article{AliMohamad2021,
  author  = {Rabei Raad Ali and Kamaruddin Malik Mohamad},
  title   = {{RX\_myKarve Carving Framework for Reassembling Complex Fragmentations of JPEG Images}},
  journal = {{Journal of King Saud University -- Computer and Information Sciences}},
  volume  = {33},
  number  = {1},
  pages   = {21--32},
  year    = {2021},
  doi     = {10.1016/j.jksuci.2018.12.007}
}

@inproceedings{Birmingham2017,
  author    = {Brandon Birmingham and Reuben A. Farrugia and Mark Vella},
  title     = {{Using Thumbnail Affinity for Fragmentation Point Detection of JPEG Files}},
  booktitle = {{IEEE EUROCON 2017 -- 17th International Conference on Smart Technologies}},
  pages     = {3--8},
  year      = {2017},
  organization = {IEEE},
  doi       = {10.1109/EUROCON.2017.8011068}
}

@article{Boiko2023OOXML,
  author  = {Maksym Boiko and Viacheslav Moskalenko},
  title   = {{Syntactical Method for Reconstructing Highly Fragmented OOXML Files}},
  journal = {{Radioelectronic and Computer Systems}},
  number  = {1},
  pages   = {166--182},
  year    = {2023},
  doi     = {10.32620/reks.2023.1.14}
}

@inproceedings{Boiko2025OOXML,
  author       = {Maksym Boiko and Viacheslav Moskalenko and Valentyna Sloma},
  title        = {{Hybrid Approach for Carving Highly Fragmented OOXML Documents}},
  booktitle    = {{2025 IEEE 13th International Conference on Intelligent Data Acquisition and Advanced Computing Systems: Technology and Applications (IDAACS)}},
  pages        = {1--5},
  year         = {2025},
  organization = {IEEE},
  doi          = {10.1109/IDAACS68557.2025.11322102}
}

@mastersthesis{Sajja2010MP3,
  author = {Abhilash Sajja},
  title  = {{Forensic Reconstruction of Fragmented Variable Bitrate MP3 Files}},
  school = {{University of New Orleans}},
  year   = {2010},
  note   = {\url{https://scholarworks.uno.edu/td/1258/}}
}

@inproceedings{Taneja2012MP3,
  author    = {Ankit Taneja and Sascha Zmudzinski and Martin Steinebach},
  title     = {{Carving and Reorganizing Fragmented MP3 Files Using Syntactic and Spectral Information}},
  booktitle = {{Audio Engineering Society 46th International Conference: Audio Forensics}},
  address   = {Denver, CO},
  year      = {2012}
}

@mastersthesis{Hendrick2025MP3,
  author = {George H. Hendrick},
  title  = {{Defragmentation of MP3 Files}},
  school = {{Louisiana State University}},
  year   = {2025},
  note   = {\url{https://repository.lsu.edu/gradschool_theses/6123/}}
}

@mastersthesis{McCain2026MoDiCo,
  author = {Joshua D. McCain},
  title  = {{Explaining and Interpreting Byte-Level File Type Classification in Deep Learning Models}},
  school = {{Louisiana State University}},
  year   = {2026},
  note   = {\url{https://repository.lsu.edu/gradschool_theses/6317/}}
}

@article{ByteSCAN2026,
  author  = {Hyeongsik Kim and Sisung Liu and Je Hyeong Hong},
  title   = {{Size-Agnostic File Fragment Classification via Fixed Byte Slicing}},
  journal = {{ETRI Journal}},
  year    = {2026},
  doi     = {10.4218/etrij.2025-0314}
}

@misc{IPED,
  author       = {{Brazilian Federal Police and contributors}},
  title        = {{IPED Digital Forensic Tool}},
  howpublished = {\url{https://github.com/sepinf-inc/IPED}},
  year         = {2026},
  note         = {Open source digital evidence processing and indexing platform with a signature-based carver. Accessed: August 3, 2026}
}

@article{Poisel2013Classification,
  author  = {Rainer Poisel and Marlies Rybnicek and Bernhard Schildendorfer and Simon Tjoa},
  title   = {{Classification and Recovery of Fragmented Multimedia Files Using the File Carving Approach}},
  journal = {{International Journal of Mobile Computing and Multimedia Communications}},
  volume  = {5},
  number  = {3},
  pages   = {50--67},
  year    = {2013},
  doi     = {10.4018/jmcmc.2013070104}
}

@article{CaseyZoun2014,
  author  = {Eoghan Casey and Rikkert Zoun},
  title   = {{Design Tradeoffs for Developing Fragmented Video Carving Tools}},
  journal = {{Digital Investigation}},
  volume  = {11},
  pages   = {S30--S39},
  year    = {2014},
  doi     = {10.1016/j.diin.2014.05.010}
}

@article{ParkLee2014DVR,
  author  = {Jungheum Park and Sangjin Lee},
  title   = {{Data Fragment Forensics for Embedded DVR Systems}},
  journal = {{Digital Investigation}},
  volume  = {11},
  number  = {3},
  pages   = {187--200},
  year    = {2014},
  doi     = {10.1016/j.diin.2014.06.001}
}

@article{Fang2020VideoCarving,
  author  = {Junbin Fang and Guikai Xi and Rong Li and Qian Chen and Puxi Lin and Sijin Li and Zoe Lin Jiang and Siu-Ming Yiu},
  title   = {{Coarse-to-Fine Two-Stage Semantic Video Carving Approach in Digital Forensics}},
  journal = {{Computers \& Security}},
  volume  = {97},
  pages   = {101942},
  year    = {2020},
  doi     = {10.1016/j.cose.2020.101942}
}

@article{AltinisikSencar2021H264,
  author  = {Enes Altinisik and Husrev Taha Sencar},
  title   = {{Automatic Generation of H.264 Parameter Sets to Recover Video File Fragments}},
  journal = {{IEEE Transactions on Information Forensics and Security}},
  volume  = {16},
  pages   = {4857--4868},
  year    = {2021},
  doi     = {10.1109/TIFS.2021.3118876}
}

@article{Bayne2018OpenForensics,
  author  = {Ethan Bayne and R. Ian Ferguson and Adam T. Sampson},
  title   = {{OpenForensics: A Digital Forensics GPU Pattern Matching Approach for the 21st Century}},
  journal = {{Digital Investigation}},
  volume  = {24},
  pages   = {S29--S37},
  year    = {2018},
  doi     = {10.1016/j.diin.2018.01.005}
}

@inproceedings{ShanmugasundaramMemon2003,
  author       = {Kulesh Shanmugasundaram and Nasir Memon},
  title        = {{Automatic Reassembly of Document Fragments via Context Based Statistical Models}},
  booktitle    = {{19th Annual Computer Security Applications Conference (ACSAC 2003)}},
  pages        = {152--159},
  year         = {2003},
  organization = {IEEE},
  doi          = {10.1109/CSAC.2003.1254320}
}

@mastersthesis{Hildebrand2026,
  author = {Samuel Hildebrand},
  title  = {{Large-Scale File Fragment Classification via Multi-View Learning}},
  school = {{Louisiana State University}},
  year   = {2026},
  note   = {\url{https://repository.lsu.edu/gradschool_theses/6315/}}
}

@mastersthesis{Waguespack2025,
  author = {Karley M. Waguespack},
  title  = {{Building a Custom U-Net-Based CNN Model for ELF Section Classification}},
  school = {{Louisiana State University}},
  year   = {2025},
  note   = {\url{https://repository.lsu.edu/gradschool_theses/6134/}}
}

\end{document}